\documentclass[twocolumn,twoside,10pt,final]{article}
\usepackage[T1]{fontenc}
\pdfoutput=1
\UseRawInputEncoding
\usepackage{fancyhdr}
\usepackage[table, dvipsnames]{xcolor}
\usepackage{framed,multirow}
\usepackage{xurl}
\usepackage{graphicx}
\graphicspath{{Figures/}}

\usepackage{booktabs}

\usepackage{geometry}
\usepackage{pdflscape}

\usepackage{stfloats}
\usepackage{soul}
\usepackage{enumitem}
\usepackage{hyperref}
\usepackage{import}
\usepackage{array}
\usepackage{dirtytalk}
\usepackage{float}
\usepackage{pgf-pie}
\usepackage{tikz}
\usepackage{pgfplots}
\usepgfplotslibrary{colorbrewer}
\pgfplotsset{width=14cm,compat=1.9}

\usetikzlibrary{fit}

\usepackage{titlesec}

\titleformat{\paragraph}
{\normalfont\normalsize\textit}{\theparagraph}{1em}{}
\titlespacing*{\paragraph}
{0pt}{3.25ex plus 1ex minus .2ex}{1.5ex plus .2ex}

\makeatletter
\newdimen\legendxshift
\newdimen\legendyshift
\newcount\legendlines
\newcommand{\bclldist}{1mm}
\newcommand{\bclegend}[3][10mm]{%
    \legendxshift=0pt\relax
    \legendyshift=0pt\relax
    \xdef\legendnodes{}%
    \foreach \lcolor/\ltext [count=\ll from 1] in {#3}%
        {\global\legendlines\ll\pgftext{\setbox0\hbox{\bcfontstyle\ltext}\ifdim\wd0>\legendxshift\global\legendxshift\wd0\fi}}%
    \@tempdima#1\@tempdima0.5\@tempdima
    \pgftext{\bcfontstyle\global\legendxshift\dimexpr\bcwidth-\legendxshift-\bclldist-\@tempdima-0.72em}
    \legendyshift\dimexpr5mm+#2\relax
    \legendyshift\legendlines\legendyshift
    \global\legendyshift\dimexpr\bcpos-2.5mm+\bclldist+\legendyshift
    \begin{scope}[shift={(\legendxshift,\legendyshift)}]
    \coordinate (lp) at (0,0);
    \foreach \lcolor/\ltext [count=\ll from 1] in {#3}%
    {
        \node[anchor=north, minimum width=#1, minimum height=5mm,fill=\lcolor] (lb\ll) at (lp) {};
        \node[anchor=west] (l\ll) at (lb\ll.east) {\bcfontstyle\ltext};
        \coordinate (lp) at ($(lp)-(0,5mm+#2)$);
        \xdef\legendnodes{\legendnodes (lb\ll)(l\ll)}
    }
    \node[draw, inner sep=\bclldist,fit=\legendnodes] (frame) {};
    \end{scope}
}
\makeatother

\pgfplotstableread[row sep=\\,col sep=&]{
    Year&General Surgery & Orthopaedic Surgery & Visceral Surgery & Neurosurgery & Dental Surgery & Reconstructive Surgery & Heart Surgery & Vascular Surgery & Maxillofacial Surgery & Robot-assisted MIS Surgery & Urological Surgery & Paediatric Surgery & Anaesthesiology & General Applications \\
        2014  & 0                 & 2                  & 0               & 0             & 0                & 0                    & 0              & 0                & 0                     & 0                           & 0                     & 0            & 0               & 0  \\
        2015  & 0                 & 1                  & 0               & 0             & 1                & 0                    & 0              & 0                & 0                     & 0                           & 0                     & 0            & 0               & 0 \\
        2016  & 0                 & 2                  & 0               & 0             & 0                & 0                    & 0              & 1                & 0                     & 0                           & 2                     & 0            & 0               & 0  \\
        2017  & 0                 & 2                  & 1               & 1             & 0                & 1                    & 2              & 1                & 0                     & 0                           & 0                     & 0            & 0               & 0  \\
        2018  & 0                 & 5                  & 0               & 2             & 1                & 1                    & 0              & 0                & 0                     & 1                           & 0                     & 0            & 0               & 0  \\
        2019  & 3                 & 6                  & 0               & 2             & 1                & 0                    & 2              & 2                & 2                     & 0                           & 0                     & 1            & 1               & 4 \\
        2020  & 2                 & 1                  & 0               & 3             & 1                & 0                    & 0              & 0                & 0                     & 1                           & 0                     & 0            & 0               & 0 \\
}\distributionSurgicalSpecialityTimelineData

\definecolor{airforceblue}{rgb}{0.36, 0.54, 0.66}
\definecolor{amethyst}{rgb}{0.6, 0.4, 0.8}
\definecolor{blush}{rgb}{0.87, 0.36, 0.51}
\definecolor{coral}{rgb}{1.0, 0.5, 0.31}
\definecolor{applegreen}{rgb}{0.55, 0.71, 0.0}
\definecolor{coquelicot}{rgb}{1.0, 0.22, 0.0}
\definecolor{atomictangerine}{rgb}{1.0, 0.6, 0.4}
\definecolor{royalblue}{rgb}{0.0, 0.14, 0.4}

\usepackage{colortbl}
\usepackage{epstopdf}
\usepackage{subcaption}
\usepackage[labelfont=bf]{caption}
\usepackage{threeparttable}

\pgfplotsset{compat=1.10}
\usetikzlibrary{decorations.pathreplacing}

\usepackage{titlesec}
\usepackage[title]{appendix}

\hypersetup{
    colorlinks=true,
    linkcolor=cyan,
    filecolor=magenta,    
    citecolor=cyan,
    urlcolor=cyan,
    pdftitle={Utility of Optical See-Through Head Mounted Displays in Augmented Reality-Assisted Surgery: A systematic review},
    pdfpagemode=FullScreen,
    }

\newcommand\blfootnote[1]{%
  \begingroup
  \renewcommand\thefootnote{}\footnote{#1}%
  \addtocounter{footnote}{-1}%
  \endgroup
}

\usepackage{natbib}

\usepackage[font=scriptsize]{caption}

\begin{document}

\twocolumn[
  \begin{@twocolumnfalse}
  \thispagestyle{empty}
\title{Utility of Optical See-Through Head Mounted Displays in Augmented Reality-Assisted Surgery: A systematic review}%
\date{}

\vspace*{-50pt}
\begin{minipage}{\textwidth}
\centering
\author{
Manuel Birlo\textsuperscript{ a},
P.J. ``Eddie'' Edwards\textsuperscript{ a}
Matthew Clarkson\textsuperscript{ a}
Danail Stoyanov\textsuperscript{ a}
}
\end{minipage}

\maketitle

\vspace*{-20pt}\hspace*{40pt}\begin{minipage}{0.9\textwidth}
\footnotesize
\textsuperscript{a} \textit{Wellcome/EPSRC Centre for Interventional and Surgical Sciences (WEISS), UCL, Charles Bell House, 43-45 Foley Street, London W1W 7TS, UK}
\end{minipage}
\vspace*{20pt}
\begin{abstract}
This article presents a systematic review of optical see-through head mounted display (OST-HMD) usage in augmented reality (AR) surgery applications from 2013 to 2020. Articles were categorised by: OST-HMD device, surgical speciality, surgical application context, visualisation content, experimental design and evaluation, accuracy and human factors of human-computer interaction. 91 articles fulfilled all inclusion criteria. Some clear trends emerge. The Microsoft HoloLens increasingly dominates the field, with orthopaedic surgery being the most popular application (28.6\%). By far the most common surgical context is surgical guidance (n=58) and segmented preoperative models dominate visualisation (n = 40). Experiments mainly involve phantoms (n = 43) or system setup (n = 21), with patient case studies ranking third (n = 19), reflecting the comparative infancy of the field. Experiments cover issues from registration to perception with very different accuracy results. Human factors emerge as significant to OST-HMD utility. Some factors are addressed by the systems proposed, such as attention shift away from the surgical site and mental mapping of 2D images to 3D patient anatomy. Other persistent human factors remain or are caused by OST-HMD solutions, including ease of use, comfort and spatial perception issues. The significant upward trend in published articles is clear, but such devices are not yet established in the operating room and clinical studies showing benefit are lacking. A focused effort addressing technical registration and perceptual factors in the lab coupled with design that incorporates human factors considerations to solve clear clinical problems should ensure that the significant current research efforts will succeed.

\end{abstract}
\vspace{20pt}\\

  \end{@twocolumnfalse}
]
\blfootnote{Corresponding author email: manuel.birlo.18@ucl.ac.uk}
\thispagestyle{empty}
\section{Introduction}
\label{section:introduction}
Augmented reality (AR) surgical guidance was originally proposed in neurosurgery over 35 years ago~\citep{Kelly_Neuro_1982,Roberts_JNeuro_1986}. The ability to view patient models directly on the surgeon's view promises numerous benefits, including better perception, ergonomics, hand-eye coordination, safety, reliability, repeatability, and ultimately improved surgical outcomes. But despite more than three decades of research, the promise of AR has not yet translated into routine clinical practice. The development of commercial optical see-through head-mounted displays (OST-HMDs) including Google Glass, Moverio and the HoloLens have led to an increasing interest in such devices for surgical guidance (see Fig.~\ref{fig:google_Scholar}). While there is occasional critical analysis \citep{carbone2020commercially}, the majority of authors tend to emphasise the great potential of AR in surgical applications. Multiple review papers for different specialities follow a similar pattern of positivity, but note that further research is needed. The lack of research demonstrating clinical benefit has been widely noted.

There is a risk that enthusiasm for the newly available OST-HMD devices may lead research along a similar path to earlier work and fail to achieve translation into routine surgery. The purpose of this paper is to provide a summary of current research in OST-AR in surgery and examine possible barriers to clinical adoption of such devices. We categorise papers according to application area, consider registration and validation methodologies as well as choice of visual content. Human factors emerge as a significant set of issues potentially limiting the applicability of AR and we provide a description of the most common issues encountered.

We believe that clinically successful AR can be achieved by clear identification of the critical points in specific surgical procedures where guidance is needed, identification of the relevant information segmented from preoperative scans and attention to human factors issues regarding the modes of visualisation and interaction. We hope that this review will be useful to clinicians and engineers collaborating on new OST-HMD AR projects and recommend consideration of human factors at an early stage. 

\section{Background}

AR systems can be realised using different types of display media such as conventional displays, projectors or head-mounted displays (HMDs) \citep{Sielhorst2008optical,Okamoto2015}. Of these three display media categories, HMDs offer the most user-friendly solution for manual tasks since the user can work both from a self-centered perspective and hands-free \citep{Condino2020}.  HMDs can be classified according to their underlying AR paradigm: Video see-through (VST) or optical see-through (OST). 

In VST systems, a video image feed is combined with superimposed computer generated images such as 3D reconstructed organs. VST systems have been adopted in surgical applications via computer displays and HMDs, and have several potential advantages including improved synchronisation between video feed and overlay as well as video processing for image segmentation or registration. Also the contrast between video feed and overlaid graphical content can be easily controlled and the real scene can be occluded by a virtual overlay. The disadvantages of VST systems include limitations in terms of video bandwidth, the risk of losing vision of the real scene in the case of system errors and geometric aberrations such as distorted spatial perception \citep{Cutolo2018}. Though video and overlay may be well synchronised, there is inevitably some delay between actual motion and perception of the motion, both real and overlaid, which can slow down surgical motion and may increase errors. \cite{guo2019online} also reported that the absence of a direct view to the real world makes surgeons nervous.

In OST-HMDs a transparent monitor displaying graphical content is located between the surgeon's line of vision and the target organ structures. This provides an unhindered view of reality, natural stereo vision capabilities with no lag or loss of resolution associated with the real scene. The downsides, however, are dynamic registration errors for the augmented view, latency when moving, static registration errors, complex calibration and unnatural perceptual issues, such as the fact that nearer virtual objects don't occlude real objects in the background~\citep{rolland1995comparison}.

\subsection{Commercial OST-HMD Devices}
Prior to 2013 research in OST-HMD relied largely on custom built devices. It is a technically difficult challenge to make such a device, incorporating miniaturised displays into a wearable headset with half-silvered mirrors enabling free view of the real scene. The optical setup to display a bright image with good contrast and resolution covering a wide field-of-view is technically hard to achieve. Only two of the papers in our review use custom devices.

The potential commercial benefits of being able to place graphical information directly overlaid on the wearer's view of the real world is a vision that has led to the development of a number of commercial devices. Google Glass, released in 2013, is a lightweight monocular AR device enabling display of information while continuing daily life. The Microsoft HoloLens, released in 2017, is a larger HMD that incorporates stereo vision, low latency room mapping and head tracking as well as gesture-based interaction using only the wearer's hands. Numerous other devices have appeared offering different levels of comfort and function (for a more detailed list of OST-HMD devices see section~\ref{section:head_mounted_displays}).

Though none of these devices were specifically designed for surgical tasks, the potential for convenient display of information to the surgeon has led to the explosion of research detailed in this review. In common with any medical intervention, the fundamental questions concern safety and efficacy.

\subsection{Safety}
Convenient overlay of information comes with inherent risks. Where the aim is that the overlay directly guides surgery, accuracy is key. Some authors are critical of OST-HMD device accuracy. \cite{Condino2020} concluded from their quantitative study that the HoloLens should not be used for high-precision manual tasks. \cite{carbone2020commercially} also conclude that OST-HMDs are unsuitable for surgical guidance, suggesting that research should focus on addressing perceptual issues that play a critical role in limiting user accuracy. \cite{Fida2018} performed a systematic review of augmented reality in open surgery and conclude that such perceptual issues limit their usage to the augmentation of simple virtual elements such as models, icons or text. 

Even accurately overlaid could distract from or hamper the surgeon's view of the patient, potentially slowing the response to critical situations such as bleeding. \cite{dilley2019perfect} propose nearby presentation of correctly oriented but not registered models. Gesture interactions with the AR view may prove difficult to combine with the manual surgical task itself~\citep{Solovjova_2019_MuC}. Cognitive overload can occur if too much extra information is presented to the surgeon at the same time~\citep{katic2015system}.

\cite{cometti2018effects} analyzed the effects of mixed reality HMDs on cognitive and physiological functions during intellectual and manual tasks that last for 90 minutes. Their experiment consisted of 12 volunteers performing  and manual tasks with and without the HoloLens while their physical and mental conditions (cognitive, cardiovascular and neuromuscular) were measured. They conclude that using the HoloLens is safe since it does not impact safety-critical human functionalities like balance and cognitive and physical fatigue. However, despite the positive outcome of the study, the authors also state that one of the prerequisites of a safe and effective usage of HMDs is that users should be receptive to the device.

While some of the technological limitations of OST-HMDs are currently being addressed, such as limited field of view and automatic eye-to-eye calibration (e.g. HoloLens 2, Microsoft Corporation, Redmond, USA), human-factor limitations remain the major hurdles that prevent the commercial success of OST-HMD AR solutions within surgical applications~\citep{Cutolo2018}.

\subsection{Efficacy}
The fundamental advantage of augmented reality surgery lies in the convenient display of graphical, image, icon or text information directly on the surgeon's view of the patient. There is no need to look away from the surgical scene or stop the operation to obtain potentially useful visual input.

When displaying guidance information, accuracy becomes a measure of system performance and the majority of the papers included in this review perform some accuracy or precision experiments. It is important to distinguish registration or tracking accuracy, which is often based on an external tracking or guidance system, from perceptual accuracy achieved by the AR system.

The ultimate test of efficacy would be improved patient outcome, but the systems reviewed are not currently at the stage of large scale clinical trials that would be needed to demonstrate patient benefit.

This review aims to give an overview of the current state of the art in OST-HMD assisted surgery by analysis of several components of the selected literature, including OST-HMD device, surgical speciality, surgical application context, surgical procedure, AR visualisations, conducted experiments and accuracy results. A special focus is given to the identification of human factors in each article. 

\begin{figure}[t!]
\centering
\resizebox{1\columnwidth}{!}{%
  \includegraphics[scale=0.7]{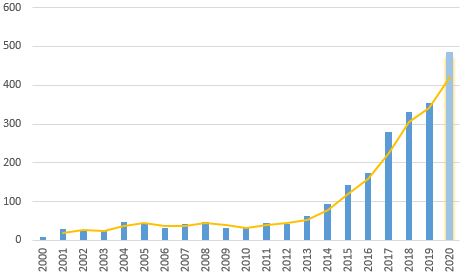}
}
\caption{Google Scholar search results for surgery "Head Mounted Display" "Augmented Reality" OR "Mixed Reality" surgery "Head Mounted Display" "Augmented Reality" OR "Mixed Reality" "optical see through" OR "Hololens" OR "Magic Leap" OR "Google Glass" in the last 20 years (2020 results have a more recent search date)}
\label{fig:google_Scholar} 
\end{figure}

\section{Methods}

\subsection{Literature search}
A systematic review was performed according to the preferred reporting items for systematic review and meta-analysis (PRISMA) guidelines~\cite{liberati2009prisma}. The literature search was conducted via a Google Scholar with the search terms [surgery ``Head Mounted Display'' ``Augmented Reality'' OR ``Mixed Reality'' surgery ``Head Mounted Display'' ``Augmented Reality'' OR ``Mixed Reality'' ``optical see through'' OR ``Hololens'' OR ``Magic Leap'' OR ``Google Glass'']. An initial Google Scholar including all articles between 2013 and 2020 was conducted on February 21, 2020. An updated Google Scholar search for 2020 only was subsequently performed on January 27, 2021. 
\subsection{Other review papers}
Since we want to analyze only original research papers, other review papers are not considered. Our search did return a number of these, however, which deserve some attention. 

A general review of all areas of AR, including medical and surgical, is provided by \cite{dey2018systematic}, who examine the usability of AR over a 10 year period. \cite{Chen_ISMAR_2017} review medical applications of mixed reality and provide a broad taxonomy. A comprehensive review of medical AR is provided by \cite{eckert2019augmented} who conclude that there is no proof of clinical effectiveness as yet. \cite{kersten2013state} give a comprehensive review using the DVV taxonomy and provide suggestions for areas that need attention, including specific overlays for important phases of the operation as well as optimisation of interaction and system validation.

We have also previously identified some of the barriers to adoption of surgical AR in general~\citep{Edwards_2021}. Existing comprehensive reviews of related surgical areas were found, including robotics~\citep{QianRoboticsReview_MRB2020} and laparoscopic surgery~\citep{bernhardt2017status}.
Orthopaedics is the dominant application area in this review and three other reviews cover this specific field well~\citep{laverdiere2019augmented,jud2020applicability,verhey2020virtual}.

It is worth noting that nearly all the review papers suggest the potential of AR in surgical applications, but cite technological hurdles to user acceptability and the lack of any clinical validation. None of these reviews cover OST-HMDs specifically, which is an increasingly popular choice. We aim to provide a critical analysis of the important characteristics of OST-HMDs, looking specifically at human factors issues, which emerged as significant area potentially limiting user acceptability of systems.

\begin{figure}[t!]
   \centering
   \resizebox{\columnwidth}{!}{%
   \includegraphics{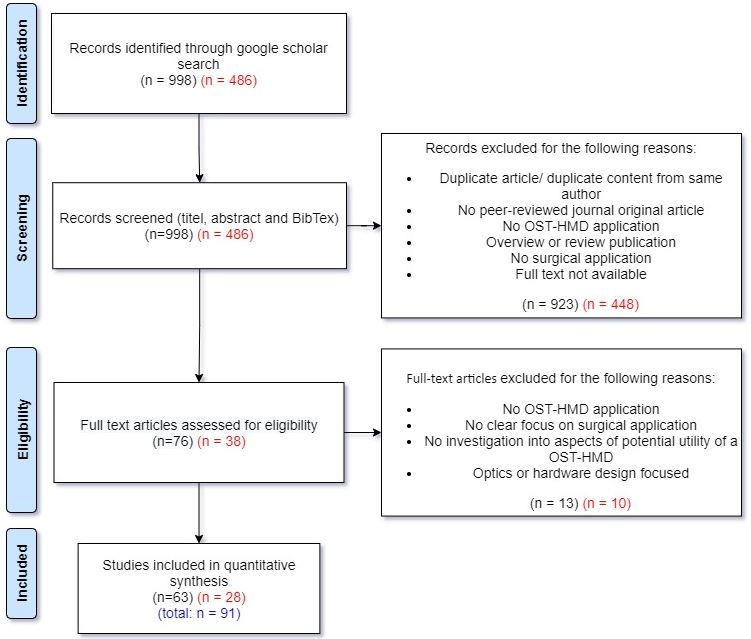}
   }
   \caption{Systematic review search strategy}
   \label{fig:review_process}
\end{figure}

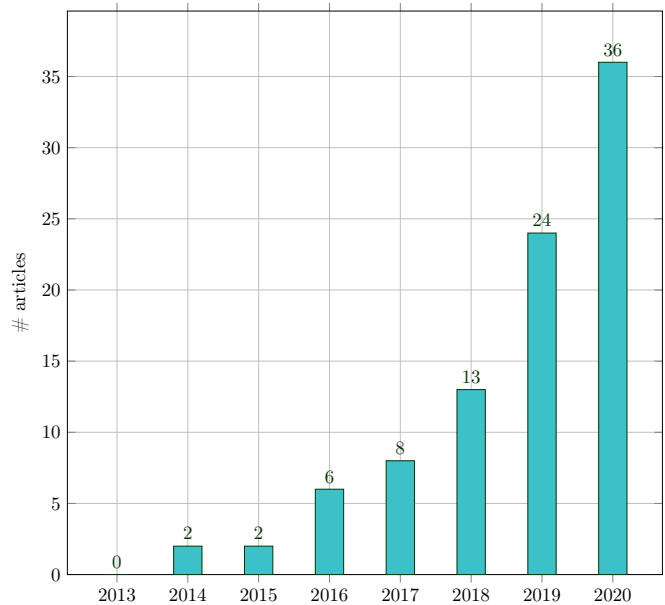
\begin{figure}[ht]
\centering
\resizebox{\columnwidth}{!}{%
\begin{tikzpicture}[scale=0.5]
\begin{axis}[
symbolic x coords={,2013, 2014, 2015, 2016, 2017, 2018, 2019, 2020, },
    ymin=0,
    ybar=-\pgfplotbarwidth,
    x=2.5*\pgfplotbarwidth,
    bar width=15pt,
	ymajorgrids,
	xmajorgrids,
	nodes near coords,
	ylabel={\# articles}
]
\addplot[green!20!black,fill=Aquamarine!80!white] coordinates{(2013,0)}; 
\addplot[green!20!black,fill=Aquamarine!80!white] coordinates{(2014, 2)}; 
\addplot[green!20!black,fill=Aquamarine!80!white] coordinates{(2015, 2)}; 
\addplot[green!20!black,fill=Aquamarine!80!white] coordinates{(2016, 6)}; 
\addplot[green!20!black,fill=Aquamarine!80!white] coordinates{(2017, 8)}; 
\addplot[green!20!black,fill=Aquamarine!80!white] coordinates{(2018, 13)}; 
\addplot[green!20!black,fill=Aquamarine!80!white] coordinates{(2019, 24)}; 
\addplot[green!20!black,fill=Aquamarine!80!white] coordinates{(2020, 36)}; 

\end{axis}
\end{tikzpicture}
}
\caption{Systematic review results overview: Annual Distribution of selected 91 studies from 2013-2020}
\label{fig:distrib_91_articles}
\end{figure}

\subsection{Literature analysis strategy}

In order to narrow down the publication year search range and focus on more recent research,
the number of publications resulting from the google scholar search terms in the last 20 years were analyzed (figure \ref{fig:google_Scholar}), which shows a steady increase from 2013, coinciding with the release of Google Glass. Due to the clear increase from that time, we chose 2013 as the starting year for our literature review.

The review process is shown in Fig.~\ref{fig:review_process} and includes the results from both the original search (February 21, 2020, numbers in black colour) and the updated search (January 21, 2021, numbers in red colour). The Google Scholar search initially resulted in 998 (486) records. In a subsequent screening phase, title, abstract and BibTex information were read to decide whether the record seems to be a relevant publication. We exclude records that are either a duplicate or contain duplicated content from the same authors compared to another record. A total of 15 (7) duplicates were excluded. During the screening the following inclusion criteria were used: The article has to 1.) be a peer-reviewed original journal article, 2.) describe an OST-HMD focused application with surgical context, 3.) is not an overview or systematic review publication (which are considered separately). Records whose full text wasn't available were excluded. 907 (441) records that didn't meet the inclusion criteria were excluded. Together with the 15 (7) excluded duplicates, a total of 923 (448) records were excluded during the screening phase, which led to 76 (38) remaining full text articles that were assessed for eligibility. Full text articles had to meet the following inclusion criteria: The article 1.) describes the usage of an OST-HMD, 2.) with a clear focus on a surgical application, 3.) investigates the potential utility of OST-HMDs in surgical settings and 4.) is neither optics nor hardware design focused. 13 (10) articles that didn't meet these inclusion criteria were excluded. 
The remaining 63 + 28 = 91 studies that met all predefined inclusion criteria form the final set of papers examined in this review. 
 
When reporting the results, the PRISMA guidelines were followed. Due to the inherent characteristics of the studies (small case series, subjective qualitative assessments, no controlled randomised trials) a meta-analysis could not be performed. Therefore, publication bias could not be reduced and should be taken into account. 

Data extracted from the included publications were
1. Clinical setting (surgical speciality, surgical application context, surgical procedure),
2. The assessed OST-HMD device, 
3. Methods (AR visualisations, Conducted experiments),
4. Key results (Accuracy), 
5. Human factors

\section{Analysis of the literature search}
This section summarises the results of the included 91 articles that were identified in the literature review process. An overview of used OST-HMD device, surgical application context and surgical procedure can be found in table~\ref{table:hmd_surgical_phase_surgical_applications}. Appendix table~\ref{table:appendix} contains details about AR visualisations, conducted experiments and accuracy results.
\begin{table*}[t]
\centering
\resizebox{\textwidth}{!}{
\begin{threeparttable}
\caption{Distribution of the included articles per surgical speciality}
    \label{table:annual_distrib_surg_speciality}
     \rowcolors{2}{white}{lightgray!20}
     \begin{scriptsize} 
     \begin{tabular}{lp{0.6\linewidth}}
        \toprule
        \multicolumn{1}{c}{\bf Surgical speciality} & \multicolumn{1}{c}{\bf Articles} \\
        \midrule 
         Orthopaedic Surgery  & \cite{armstrong2014heads} \cite{ponce2014emerging} \cite{chen2015development}
         \cite{wang2016precision} \cite{stewart2016wearable} 
        \cite{hiranaka2017augmented} \cite{jalaliniya2017wearable} \cite{unberath2018augmented}  \cite{el-hariri2018augmented}
         \cite{deib2018image}  \cite{Andres2018}  \cite{condino2018build}  \cite{gibby2019head} \cite{deOliveira2019} \cite{aaskov2019x} \cite{liebmann2019pedicle} \cite{fotouhi2019interactive} \cite{fotouhi2019co}  
        \cite{pietruski2020supporting}   
        \cite{laguna2020assessing} \cite{gibby2020use} \cite{gu2020feasibility} \cite{viehofer2020augmented} \cite{dennler2020augmented} \cite{kriechling2020augmented} \cite{matsukawa2020smart} \\
        
        General Surgery & \cite{lin2018holoneedle}  \cite{wu2018augmented} \cite{mahmood2018augmented} \cite{rojas2019surgical} \cite{li2019mixed} \cite{pelanis2020use} \cite{zhou2020surgical} \cite{al2020effectiveness} 
          \cite{jud2020applicability} \cite{galati2020experimental} \cite{li2020clinical} 
         \\
        Neurosurgery & \cite{yoon2017technical} \cite{karmonik2018augmented} \cite{frantz2018augmenting}  \cite{nguyen2020augmented} \cite{zhang2019preliminary} \cite{heinrich2019holoinjection} \cite{baum2020augmented} \cite{liounakos2020head} 
        \cite{sun2020high} \\ 
        
        Vascular Surgery & \cite{kaneko2016ultrasound} \cite{kuhlemann2017towards} \cite{zhou2019design} \cite{rynio2019holographically} \cite{rojas2020system}  
        \cite{park2020three} \cite{mendes2020pinata}  
        \\
        
        General surgical applications  & \cite{meulstee2019toward}  \cite{chien2019hololens} \cite{boillat2019increasing} \cite{guo2019online} 
        \cite{dallas2020comparing} \cite{luzon2020value} \cite{cartucho2020multimodal} \cite{kumar2020use} 
        \\   
        
        Dental Surgery   &   \cite{katic2015system} \cite{liebert2016novel} \cite{song2018endodontic} \cite{zhou2019towards}  
        \cite{zafar2020evaluation} 
        \\ 
        
        Heart Surgery & \cite{li2017human} \cite{zou2017coronary} \cite{brun2019mixed} \cite{liu2019augmented}\\ 
        
        Otolaryngology - head and neck surgery & \cite{rojas2020evaluation} \cite{gnanasegaram2020evaluating} \cite{scherl2020augmented} \cite{creighton2020early} 
        \\
        
        Reconstructive Surgery  & \cite{mitsuno2017intraoperative}  \cite{pratt2018through} 
        \cite{jiang2020hololens} 
        \\
        
        Maxillofacial Surgery  & \cite{pietruski2019supporting} \cite{pepe2019marker} 
        \cite{sun2020fast} 
        \\

        Urological surgery & \cite{borgmann2016feasibility} \cite{dickey2016augmented}
        \cite{schoeb2020mixed} 
        \\
        
         Robot-assisted surgery & \cite{qian2018arssist} \cite{fotouhi2020reflective} \\ 
         
        Paediatric surgery  & \cite{wellens2019comparison} 
        \cite{fitski2020mri} 
        \\
        Visceral Surgery & \cite{sauer2017mixed} \\ 
         
        Interventional Oncology & \cite{li2020smartphone} \\ 
        
        Laparoscopic Surgery & \cite{zorzal2020laparoscopy} \\ 
        
        Anaesthesiology & \cite{schlosser2019exploratory} \\ 
        \bottomrule 
     \end{tabular}
     \end{scriptsize}
\end{threeparttable}
}
\end{table*}

\subsection{Annual distribution of selected articles}

Fig. \ref{fig:distrib_91_articles} shows the annual distribution of the 91 studies during 2013-2020. There were no articles in the year 2013 that fulfilled the inclusion criteria. Starting from 2014 there has been a steady increase in the number of publications. The increasing trend tends to be related to the release of major OST-HMDs like Google Glass and Microsoft HoloLens, and will be discussed in more detail in section~\ref{section:head_mounted_displays}.

\subsection{Surgical speciality}

\begin{figure}[ht!]
    \centering
    \resizebox{\columnwidth}{!}{%
      \includegraphics[width=0.45\textwidth]{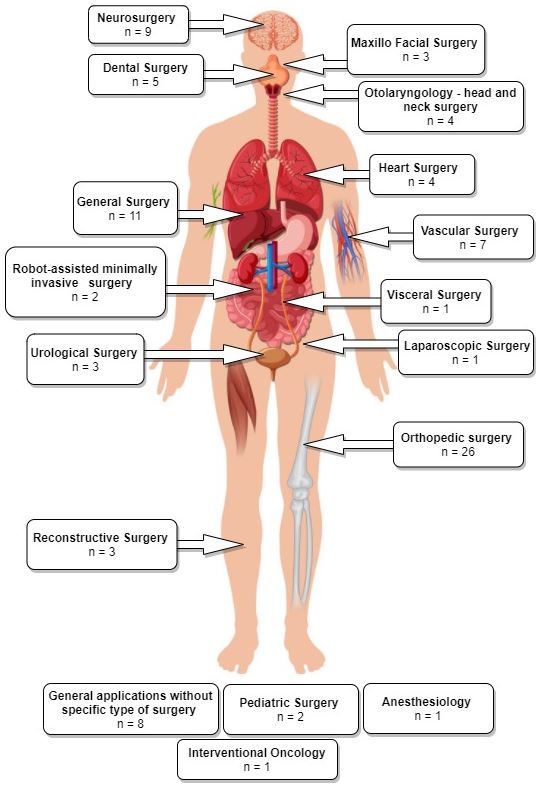}
      }
       \caption{Graphical illustration of included articles grouped by surgical speciality and placed at respective human body regions}
       \label{fig:human_body_overview}
\end{figure}

\begin{figure}[ht!]
    \centering
    \resizebox{\columnwidth}{!}{
      \begin{tikzpicture}[object/.style={thin,double,<->}]
       \pie{
             28.6/Orthopaedic Surgery, 
             12.1/General Surgery, 
             9.9/Neurosurgery, 
             8.8/General Applications, 
             7.7/Vascular Surgery, 
             5.5/Dental Surgery, 
             4.4/Heart Surgery, 
             4.4/Otolaryngology, 
             3.3/Reconstructive Surgery, 
             3.3/Urological Surgery, 
             3.3/Maxillofacial Surgery, 
             2.2/Paediatric Surgery, 
             2.2/Robot-assisted Surgery, 
             4.4/Other Surgical Specialities 
           }
    \end{tikzpicture}
    }
       \caption{Pie chart showing the distribution of the included 91 papers among the identified surgical specialities}
       \label{fig:pie_chart}
\end{figure}
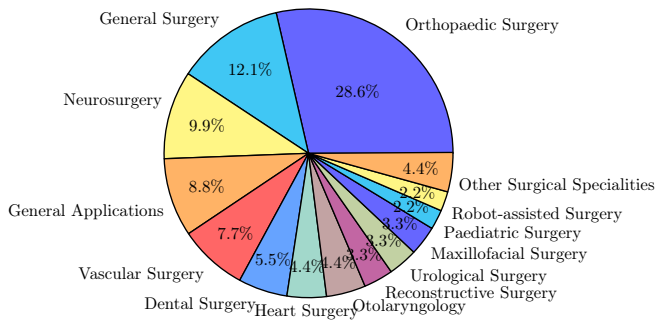

We found that OST-HMDs have been applied in a variety of surgical specialities. 
Fig. \ref{fig:human_body_overview} shows a graphical illustration of all articles grouped into their surgical speciality and placed at their respective body region. Fig. \ref{fig:pie_chart} shows the proportion of publications for each surgical speciality. Orthopaedic surgery dominates (28.6\%, n = 26), perhaps since proximity to bone requires only rigid registration and somewhat lower accuracy is required compared to applications such as neurosurgery. 
General surgery, neurosurgery, applications without a concrete surgical speciality and vascular surgery follow with more than five articles each. Dental surgery is represented with five articles, followed by heart surgery and Otolaryngology (n = 4 each). Other surgical specialities include reconstructive surgery, urology and maxillofacial surgery (n = 3 each). A few attempts have been made to explore potential benefits of OST-HMDs in robot-assisted surgery and paediatric surgery (n = 2 each). Interventional oncology, laparoscopic surgery, visceral surgery and anaesthesiology are represented with one article.
Specific articles per surgical speciality are detailed in table~\ref{table:annual_distrib_surg_speciality}. While orthopaedics still dominates, other applications, including general, vascular and neurosurgery, are increasingly represented in the latter half of the survey period as interest in AR applications spreads to other surgical fields.

\subsection{Optical see-through head-mounted displays (OST-HMDs)}
\label{section:head_mounted_displays}

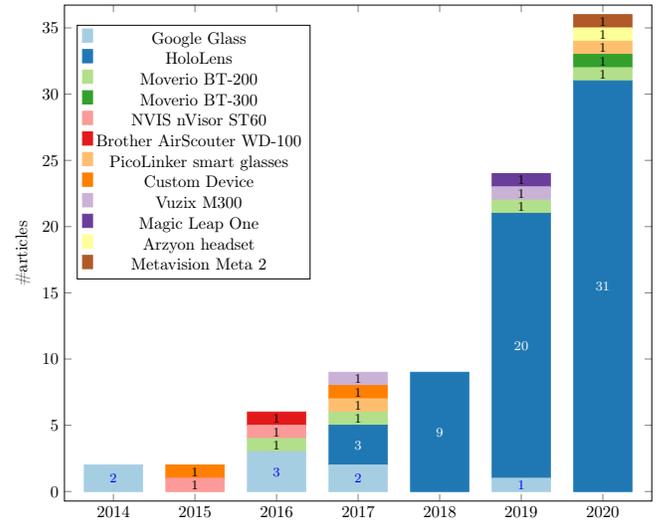
\begin{figure}[ht!]
\resizebox{\columnwidth}{!}{%
\usepgfplotslibrary{colorbrewer}
\begin{tikzpicture}[scale=0.4]
\begin{axis}[
    enlarge y limits=0.02,
    xtick=data,
	ybar stacked,
	bar width=35pt,
	nodes near coords,
	every node near coord/.append style={font=\footnotesize},
    clip=false,
    legend style={at={(0.22, 0.96)}, anchor=north,legend columns=1},
    ylabel={\#articles},
    symbolic x coords={2014, 2015, 2016, 2017, 2018, 2019, 2020},
    xtick=data,
    cycle list/Paired,
    every axis plot/.append style={fill}
    ]
\addplot +[ybar, text=blue] plot coordinates {(2014, 2) (2015, 0) (2016, 3) (2017, 2) (2018, 0) (2019, 1) (2020, 0) }; 
\addplot +[ybar, text=white] plot coordinates {(2014, 0) (2015, 0) (2016, 0) (2017, 3) (2018, 9) (2019, 20) (2020, 31) }; 
\addplot +[ybar, text=black] plot coordinates {(2014, 0) (2015, 0) (2016, 1) (2017, 1) (2018, 0) (2019, 1) (2020, 1) }; 
\addplot +[ybar, text=black] plot coordinates {(2014, 0) (2015, 0) (2016, 0) (2017, 0) (2018, 0) (2019, 0) (2020, 1) }; 
\addplot +[ybar, text=black] plot coordinates {(2014, 0) (2015, 1) (2016, 1) (2017, 0) (2018, 0) (2019, 0) (2020, 0) }; 
\addplot +[ybar, text=black] plot coordinates {(2014, 0) (2015, 0) (2016, 1) (2017, 0) (2018, 0) (2019, 0) (2020, 0) }; 
\addplot +[ybar, text=black] plot coordinates {(2014, 0) (2015, 0) (2016, 0) (2017, 1) (2018, 0) (2019, 0) (2020, 1) }; 
\addplot  +[ybar, text=black] plot coordinates {(2014, 0) (2015, 1) (2016, 0) (2017, 1) (2018, 0) (2019, 0) (2020, 0) }; 
\addplot +[ybar, text=black] plot coordinates {(2014, 0) (2015, 0) (2016, 0) (2017, 1) (2018, 0) (2019, 1) (2020, 0) }; 
\addplot +[ybar, text=black] plot coordinates {(2014, 0) (2015, 0) (2016, 0) (2017, 0) (2018, 0) (2019, 1) (2020, 0) }; 
\addplot +[ybar, text=black] plot coordinates {(2014, 0) (2015, 0) (2016, 0) (2017, 0) (2018, 0) (2019, 00) (2020, 1) }; 
\addplot +[ybar, text=black] plot coordinates {(2014, 0) (2015, 0) (2016, 0) (2017, 0) (2018, 0) (2019, 00) (2020, 1) }; 
\legend{Google Glass, HoloLens, Moverio BT-200, Moverio BT-300, NVIS nVisor ST60, Brother AirScouter WD-100, 
        PicoLinker smart glasses, Custom Device, Vuzix M300, Magic Leap One, Arzyon headset, Metavision Meta 2}
\end{axis}
\end{tikzpicture}
}
\caption{Annual Distribution of articles by \textbf{OST-HMD} device from 2014-2020}
\label{fig:annual_distr_OST-HMDs}
\end{figure}

Fig. \ref{fig:annual_distr_OST-HMDs} depicts the annual distribution by OST-HMD between 2014-2000. Google Glass, the device with the second highest number of articles (n = 8), dominates the distribution in 2014, but interest decreases from 2015 to 2017, perhaps due to diminishing support from Google. The Microsoft HoloLens was released in 2016 and has dominated the field of OST-HMD assisted surgery since then, with a steady increase in papers from 2017 and accounting for the majority of articles (n = 66). Following HoloLens and Google Glass, the Moverio BT-200, which was released in 2014, has the third highest number of articles (n = 4) and was used once in 2016, 2017, 2019 and 2020. Its successor, the Moverio BT-300, was released in late 2016 and has only one application in 2020. The Magic Leap One, released in 2018 and attracting huge initial investment, has not established itself in OST-HMD assisted surgery, generating only one article in 2019. Other devices include the NVIS nVisor ST (n = 2), Vuzix M300, Brother AirScouter WD-100, Aryzon headset, Metavision Meta 2 and PicoLinker (n = 1 each).

To summarise, the HoloLens clearly dominates the field, but there is interest in other devices such as Moverio BT. This is a rapidly developing field at present and we can expect further devices to appear on the market in the next few years.

\section{Surgical application context}
\label{section:surgical_context}
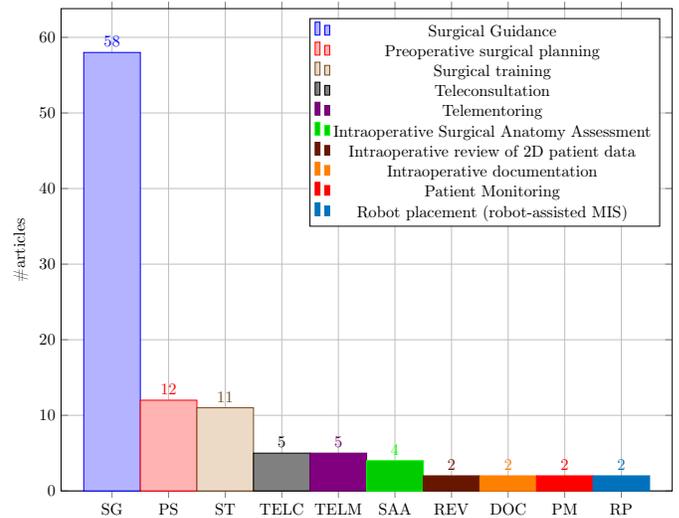
\begin{figure}[ht!]
\resizebox{\columnwidth}{!}{%
\begin{tikzpicture}
\begin{axis}[
symbolic x coords={, SG, PS, ST, TELC, TELM, SAA, REV, DOC, PM, RP, },
    ymin=0,
    ybar=-\pgfplotbarwidth,
    x=\pgfplotbarwidth,
    bar width=35pt,
	ymajorgrids,
	xmajorgrids,
	nodes near coords,
	ylabel={\#articles}
]
\addplot coordinates{(SG,58)}; 
\addplot coordinates{(PS,12)}; 
\addplot coordinates{(ST,11)}; 
\addplot coordinates{(TELC,5)}; 
\addplot coordinates{(TELM,5)}; 
\addplot coordinates{(SAA, 4)}; 
\addplot[Sepia,fill] coordinates{(REV,2)}; 
\addplot[orange,fill] coordinates{(DOC,2)}; 
\addplot[red,fill] coordinates{(PM,2)}; 
\addplot[RoyalBlue,fill] coordinates{(RP, 2)}; 
\legend{Surgical Guidance, Preoperative surgical planning, Surgical training, 
Teleconsultation, Telementoring, Intraoperative Surgical Anatomy Assessment,  Intraoperative review of 2D patient data, 
Intraoperative documentation, Patient Monitoring, 
Robot placement (robot-assisted MIS)}
\end{axis}
\end{tikzpicture}
}
\caption{Distribution of articles by \textbf{surgical application context}}
\label{fig:distrib_surgical_context}
\end{figure}

Surgical application contexts define how OST-HMD assistance is intended to improve surgical practice. Fig.~\ref{fig:distrib_surgical_context} shows the distribution of all identified contexts. Surgical guidance is by far the most popular (n=58), followed by
preoperative surgical planning (n=12) and surgical training (n = 11) then Teleconsultation and telementoring (n = 5 each). Four articles were included where the surgeon views a 3D patient anatomy hologram to aid clinical decision making rather than intraoperative guidance, which we called \textit{intraoperative surgical anatomy assessment}.
The remaining applications that have been identified are intraoperative review of preoperative 2D imaging and/or patient records, intraoperative documentation, patient monitoring and preparation of robot-assisted MIS (n = 2 each).
We expand on some of the surgical application contexts and respective articles 
in the following subsections. 

\subsection{Surgical guidance}

\begin{figure}[ht!]
\resizebox{\columnwidth}{!}{%
\begin{tikzpicture}
\begin{axis}[
symbolic x coords={,NI, SI, CI, KWI, DTG, TP, SSN, RP, SP, EG, PN, CA, IO, DG, AI, },
    ymin=0,
    ymax=14,
    ybar=-\pgfplotbarwidth,
    x=\pgfplotbarwidth,
    bar width=30pt,
	ymajorgrids,
	xmajorgrids,
	nodes near coords,
	legend style={at={(0.5, 0.98)}, anchor=north,legend columns=3},
	ylabel={\#articles}
]

\addplot[airforceblue, fill] coordinates{(NI, 8)}; 
\addplot[airforceblue!80, fill] coordinates{(SI, 7)}; 
\addplot[airforceblue!60, fill] coordinates{(CI, 4)}; 
\addplot [airforceblue!40,fill] coordinates{(KWI, 4)}; 
\addplot[airforceblue!20,fill] coordinates{(DTG, 3)}; 

\addplot[amethyst, fill] coordinates{(TP, 7)}; 
\addplot[amethyst!80, fill] coordinates{(SSN, 2)}; 
\addplot[amethyst!60,fill] coordinates{(RP, 2)}; 
\addplot[amethyst!40,fill] coordinates{(SP, 1)}; 

\addplot[blush!80,fill] coordinates{(EG, 4)}; 
\addplot[blush!60,fill] coordinates{(PN, 1)}; 
\addplot[blush!40,fill] coordinates{(CA, 1)}; 

\addplot[coral, fill] coordinates{(IO, 10)}; 
\addplot[coral!80,fill] coordinates{(DG, 4)}; 
\addplot[coral!60,fill] coordinates{(AI, 1)}; 

\legend{Needle insertion (NI), Screw insertion (SI), Catheter insertion (CI), K-Wire insertion (KWI), Drill trajectory guidance (DTG), Surgical tool placement (TP), Surgical saw navigation (SSN), Robot placement (RP), 
Stent-graft placement (SP), MIS endoscopy guidance (EG), Imaging probe navigation (PN), C-arm Positioning Guidance (CA), Image overlay for navigation (IO),  Dissection guidance (DG), 
Anatomy Identification (AI)}
\end{axis}
\end{tikzpicture}
}
\caption{Surgical guidance applications: Distribution of the subset of final 91 articles (n = 59) by applications of \textbf{surgical guidance}, grouped into the four categories \textcolor{airforceblue}{1. navigation of a linear path}, \textcolor{amethyst}{2. navigation of surgical tools or equipment}, \textcolor{blush}{3. navigation of an imaging device}, \textcolor{coral}{4. general guidance to help spatial awareness not associated with a specific task} }
\label{fig:surgical_guidance_applications}
\end{figure}
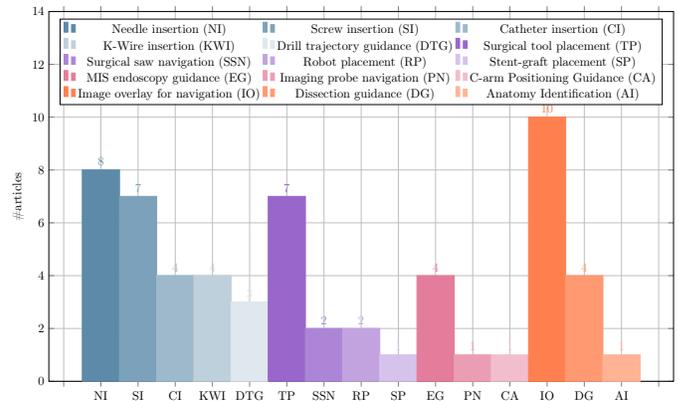

\begin{figure*}[ht]
\centering
\resizebox{\textwidth}{!}{%
\begin{tabular}{ccc}
\includegraphics[height=0.4\linewidth]{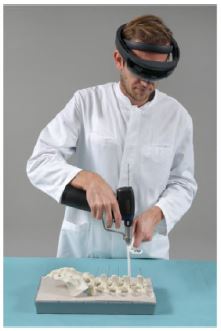}  &
\includegraphics[height=0.4\linewidth]{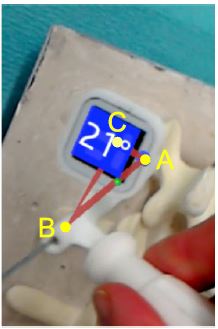} &
\includegraphics[height=0.4\linewidth]{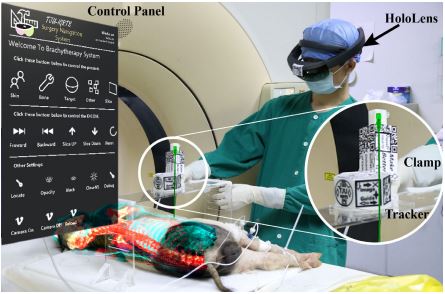}
\\
(a) & (b) & (c)\\
\end{tabular}
}
\caption{Guided screw insertion and needle insertion examples. (a) A surgeon uses a custom-made navigation device in an experimental setup (b).
Augmented drill entry points (shown in blue) are used to start the navigation. During the guided drill procedure, the 3D angle between current and targeted
screw trajectory and their deviation angle are displayed. \textit{(source: \cite{liebmann2019pedicle} Fig. 5b and 5d)}. (c) Mixed reality needle insertion navigation system for low dose-rate (LDR) brachytherapy \textit{(source: \cite{zhou2019design}) Fig. 1}.}
\label{fig:liebmann2019pedicle_fig5_b__d_and_zhou2019design_fig1}   
\end{figure*}

We use the definition of surgical guidance or image-guided surgery from \cite{cleary2010image}: a medical procedure in which a surgeon uses computer-based virtual pre- or intraoperative image overlays to visualise and target patient anatomy. They also state that an image-guided intervention includes registration and tracking methods, but we also consider an OST-HMD based solution to be of the category image guidance if it uses registered holographic image overlays without tracking if these overlays support a clinician in visualizing and targeting the surgical site.

Since this broad definition covers over half the included papers, we further split OST-HMD assisted surgical guidance into different applications, whose distribution is presented in Fig.~\ref{fig:surgical_guidance_applications}. 
General image overlay for navigation systems (n = 10) overlay a registered 3D anatomy model in order to provide surgical guidance, including applications in neuronavigation~\citep{frantz2018augmenting,nguyen2020augmented}, orthopaedic procedures \citep{deOliveira2019},
algorithm-focused registration approaches \citep{wu2018augmented,aaskov2019x,chien2019hololens} and maxillo-facial tumor resection \citep{pepe2019marker}.

Needle insertion (n = 8) has emerged as an application since 2018, mostly using the HoloLens, and was investigated in percutaneous spine procedures \citep{deib2018image}, needle biopsy \citep{lin2018holoneedle}, thoracoabdominal brachytherapy \citep{zhou2019design,zhou2020surgical} and needle-based spinal interventions \citep{heinrich2019holoinjection}. \cite{zhou2019design} presented a mixed reality based needle insertion navigation system for low-dose-rate brachytherapy that was tested in animal (Fig. \ref{fig:liebmann2019pedicle_fig5_b__d_and_zhou2019design_fig1} (c)) and phantom experiments. Reported benefits of this needle insertion approach include clinically acceptable needle insertion accuracy and a reduction of the number of required CT scans.

Tool placement examples (n = 7) include investigated attentiveness to the surgical field during navigation \citep{stewart2016wearable}, a first assistant's task performance during robot-assisted laparoscopic surgery based tool manipulation \citep{qian2018arssist}, bone localisation \citep{el-hariri2018augmented}, an optical navigation concept \citep{meulstee2019toward}, liver tumor puncture \citep{li2019mixed}, craniotomy assistance \citep{zhang2019preliminary} and percutaneous orthopaedic treatments \citep{fotouhi2019co}.
 
OST-HMD assisted screw insertion (n = 7) has been explored with different holographic visualisations. \cite{yoon2017technical} presented an application for pedicle screw placement in spine instrumentation that streamed 2D neuronavigation images onto a Google Glass. Surgeons reported an overall positive AR-experience. \cite{liebmann2019pedicle} developed a HoloLens pedicle screw placement approach for spinal fusion surgery that uses holopgraphic 3D angles between current and targeted screw trajectory, using deviation in angle to guide the surgeon (Fig.~\ref{fig:liebmann2019pedicle_fig5_b__d_and_zhou2019design_fig1} (a) and (b)). The reported results of a lumbar spine phantom experiment indicate a promising screw insertion accuracy with the caveat that surrounding tissue was not taken into account. Other articles describing pedicle screw insertion include \cite{yoon2017technical} and \cite{gibby2019head}. Percutaneous implantation of sacroiliac joint screws is presented in \cite{chen2015development} and \cite{wang2016precision}.

\begin{figure*}[ht!]
\centering
\resizebox{\textwidth}{!}{%
\begin{tabular}{cc}
\includegraphics[height=0.4\linewidth]{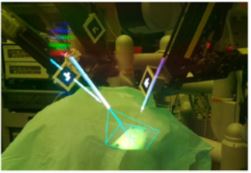} &
\includegraphics[height=0.4\linewidth]{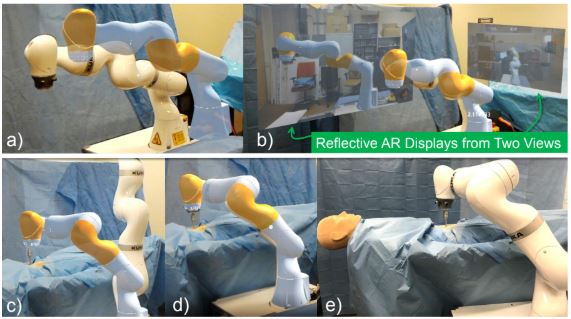} \\
(a) & (b) \\
\end{tabular}
}
\caption{(a) Robotic instrument placement and endoscopy guidance: Navigation aids for the first assistant: Real-time renderings of a robotic endoscope and robotic instruments that are superimposed on their physical counterparts. In addition,  endoscopy guidance is realised via an endoscopy visualisation being registered with a viewing frustrum \textit{(source: Fig. 4 (f) of \cite{qian2018arssist})} (b) Robot placement: Reflective-AR Display aided alignment between a real robot arm and its virtual counterpart and subsequent robot placement to its intended position in preparation for robotic surgery \textit{(source. Fig. 4 of \cite{fotouhi2020reflective})}}
\label{fig:qian2018arssist_fig4f_and_fotouhi2020reflective_fig4}
\end{figure*}

Catheter insertion (n=4) also has to deal with the manipulation of flexible structures and has been applied to US-guided central venous catheterisation \citep{kaneko2016ultrasound}, radiaton-free endovascular stenting of aortic aneurysm \citep{kuhlemann2017towards} and transcatheter procedures for structural heart disease \citep{liu2019augmented}. K-wire insertion in orthopaedic procedures (n = 4) was addressed by experiments investigating fluoroscopy controlled wire insertion into femur \citep{hiranaka2017augmented}, percutaneous orthopaedic surgical procedures \citep{Andres2018} and C-arm fluoroscopy guidance \citep{fotouhi2019interactive}. The exploration of potential benefits of holographic camera views for endoscopy guidance (n = 4) has been conducted in first assistant support in robot-assisted laparoscopic surgery \citep{qian2018arssist} (Fig. \ref{fig:qian2018arssist_fig4f_and_fotouhi2020reflective_fig4} (a)), percutaneous endoscopic lumbar discectomy \citep{liounakos2020head} and ureteroscopy \citep{al2020effectiveness}.

Drill trajectory guidance (n = 3) explores potential advantages of holographic guidance information such as drill angle and deviation between actual and planned drill path and has been used in dental implant surgery \citep{katic2015system} and endodontic treatments \citep{song2018endodontic}. Surgical saw navigation using holographic cutting guides (n = 2) was presented in mandibular resection \citep{pietruski2019supporting} and free ﬁbula ﬂap harvest \citep{pietruski2020supporting}.

\begin{figure*}[ht!]
\centering
\resizebox{\textwidth}{!}{
\begin{tabular}{cc}
\includegraphics[height=0.4\textwidth]{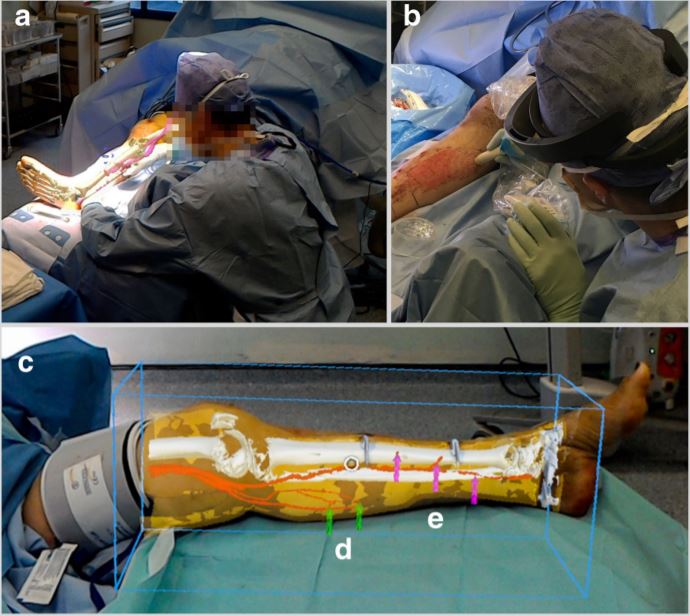} &
\includegraphics[height=0.4\textwidth]{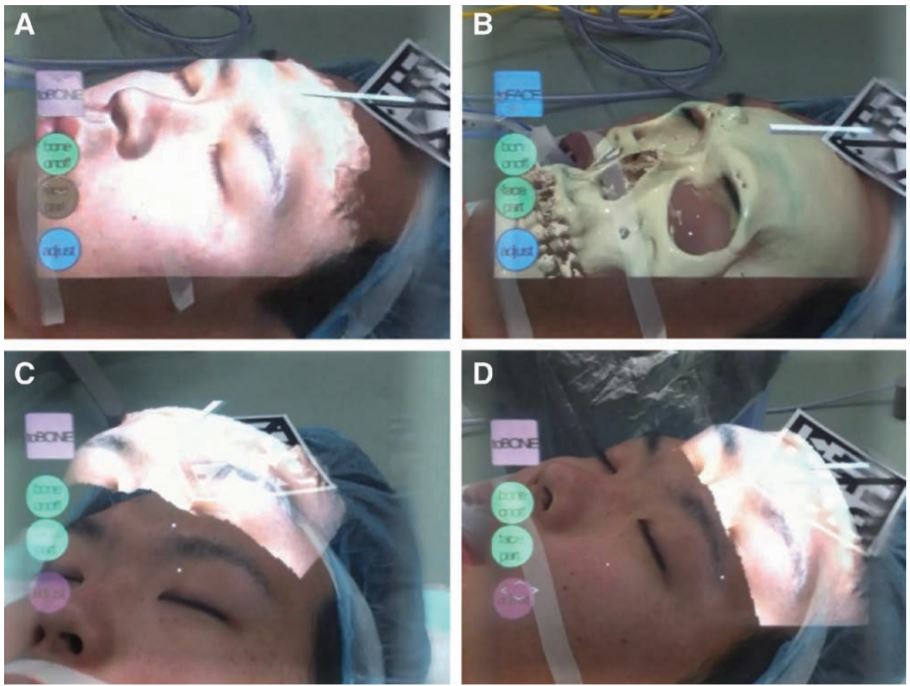} \\
(a) & (b) \\
\end{tabular}
}
\caption{(a) Dissection Guidance example in reconstructive surgery: HoloLens based identification of vascular pedunculated flaps: a CTA-based 3D model of a female patient's leg consisting of segmented skin, bone, bone, vessels and vascular perforators lower leg is superimposed on the patient anatomy. The surgeon confirms perforator location with audible Doppler ultrasonography \textit{(source: Fig. 3 of \cite{pratt2018through})} (b) Surgical Anatomy Assessment example in plastic surgery: AR views of the Moverio BT-200 smart glasses showing a patient with osteoma and holographic facial anatomy (face surface and facial bones including the osteoma) superimposed onto a patient's face \textit{(source: Fig. 8 of \cite{mitsuno2017intraoperative})}}
\label{fig:pratt2018through_fig3_and_mitsuno2017intraoperative_fig8}
\end{figure*}

In addition to surgeons themselves, other clinical staff in the operating theatre can benefit from OST-HMD assitance. In minimally invasive robotic surgery it is usually the first assistant's responsibility to set up the robot arms prior to intraoperative robot control conducted by a surgeon. We identified 2 articles that present HoloLens applications aiming to support the first assistant during robot-assisted surgery: 1.) \cite{qian2018arssist} robotic instrument placement in laparoscopic surgery from  (Fig. \ref{fig:qian2018arssist_fig4f_and_fotouhi2020reflective_fig4} (a)) and 2.) full robot arm placement in minimally invasive gastrectomy (abdominal surgery) from \cite{fotouhi2020reflective} (Fig. \ref{fig:qian2018arssist_fig4f_and_fotouhi2020reflective_fig4} (b)). 
The remaining applications of surgical guidance cover topics such as stent-graft placement in endovascular aortic repair \citep{rynio2019holographically}, imaging probe navigation for tooth decay management \citep{zhou2019towards}, C-arm positioning guidance in percutaneous orthopaedic procedures \citep{unberath2018augmented}, identification of spinal anatomy underneath the skin \citep{aaskov2019x} and dissection guidance for vascular pedunculated flaps of the lower extremities presented by \cite{pratt2018through} (Fig. \ref{fig:pratt2018through_fig3_and_mitsuno2017intraoperative_fig8} (a)). A HoloLens based mixed reality approach was decised in which the surgeon has to manually register a CTA-based 3D model of a patient's leg to the respective patient anatomy using HoloLens hand gesture and voice command interaction. After a surgical patient case study, surgeons confirmed that this mixed reality solution is more reliable and less time consuming than audible Doppler ultrasound which is the conventional non-AR method.  

With the main research focus being image-guidance, it is very important to consider safety and accuracy in such systems. Overconfidence in the accuracy of guidance or visual clutter of the viewed scene may lead to an increase in surgical errors and a careful balance needs to be struck to provide useful information rather than cognitive overload.

\subsection{Other surgical application contexts}
\label{section:other_surgical_contexts}
\begin{figure*}[ht!]
\centering
\resizebox{\textwidth}{!}{
\begin{tabular}{cc}
\includegraphics[height=0.4\textwidth]{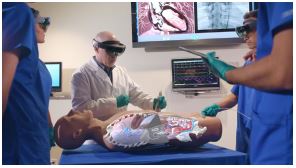}  &
\includegraphics[height=0.4\textwidth]{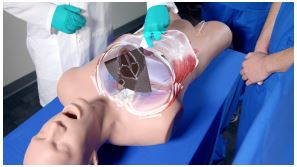}  
 
\\
(a) & (b)\\
\end{tabular}
}
\caption{Surgical Training Example Application: Ultrasound Education. (a) Multiple users can see holographic anatomical cross sections mapped on a patient simulator and the ultrasound scan plane. (b) Holograhic subcostal four-chamber view coming out of the simulator probe. \textit{Source: Fig. 3 and 7 of \cite{mahmood2018augmented}}}
\label{fig:mahmood2018augmented_3_and_7}       
\end{figure*}

Preoperative planning applications from \cite{zou2017coronary} and \cite{li2017human} addressed human-computer interaction issues of conventional approaches in preoperative diagnosis of coronary heart disease that lead to inaccurate diagnosis results and propose a hand gesture based interactive holographic diagnosis system aiming to provide a natural and intuitive interaction. \cite{karmonik2018augmented} used holographic 3D vascular structures to improve the extraction and communication of complex MRI image data  in the context of aneurysm rupture prediction. \cite{pelanis2020use} addressed planning of liver resection surgery and found that 3D holographic liver anatomy visualisations improve the user's spatial understanding.
Other articles that were categorised as preoperative surgical planning investigate potential planning improvements for repair of complex congenital heart disease \citep{brun2019mixed} and preoperative anatomy assessment for nephron-sparing surgery \citep{wellens2019comparison}.

Benefits of OST-HMD AR during surgical training have been explored for preoperative diagnosis and planning of coronary heart disease \citep{li2017human}, intraoperative surgical tool guidance during hip arthroplasty simulation \citep{condino2018build}, neurosurgical burr hole localisation \citep{baum2020augmented} and transesophageal echocardiography examination from \cite{mahmood2018augmented} shown in Fig. \ref{fig:mahmood2018augmented_3_and_7}.

\begin{figure*}[ht!]
\centering
\resizebox{\textwidth}{!}{%
\begin{tabular}{cc}
\includegraphics[height=0.45\textwidth]{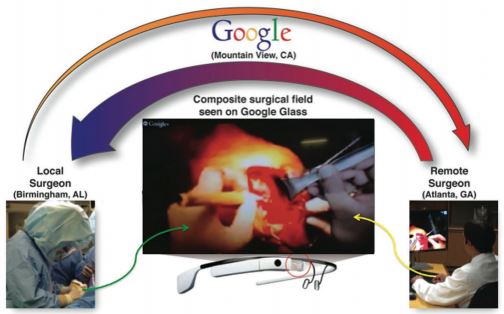}  &
\includegraphics[height=0.45\textwidth]{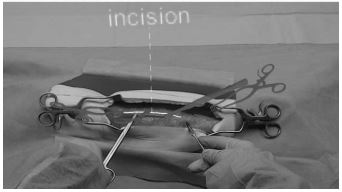}  
 
\\
(a) & (b)\\
\end{tabular}
}
\caption{Telementoring Applications. (a) Overview of a Google Glass systeming using a composite surgical field \textit{Source: Fig. 3 of \cite{ponce2014emerging}}. (b) First-person view of HoloLens-based holographic instructions consisting of 3D models and 3D lines \textit{Source: Fig. 2 of \cite{rojas2019surgical}}.}
\label{fig:ponce2014emerging_fig3_and_rojas2019surgical_fig2}       
\end{figure*}

\begin{figure*}[ht!]
\centering
\resizebox{\textwidth}{!}{%
\begin{tabular}{cc}
 \includegraphics[height=0.45\textwidth]{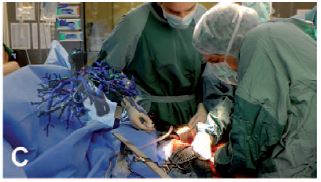} &   \includegraphics[height=0.45\textwidth]{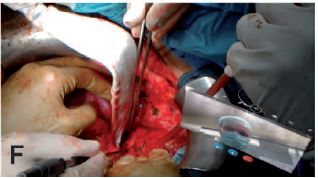} \\
(a) & (b) \\
\end{tabular}
}
\caption{Surgical Anatomy Assessment and Teleconsultation Applications in Visceral Surgery: (a) Intraoperative visualisation of a preoperative model of the vascular anatomy of the cranio-ventral liver and tumor to be dissected. (b) Intraoperative tele-consulting: real-time video communication with a remote surgeon \emph{(Source: Figure 3 (C and F) of \cite{sauer2017mixed}}) }
\label{fig:sauer2017mixed_fig3}
\end{figure*}

Telementoring also belongs to the broader scope of surgical training, but involves a surgical trainee being mentored by an expert surgeon during a surgical procedure rather than training outside the operating room. \cite{ponce2014emerging} used a Google Glass based mentoring system for shoulder arthroplasty  (Fig. \ref{fig:ponce2014emerging_fig3_and_rojas2019surgical_fig2} (a)). The student surgeon and teacher surgeon can both see a composite surgical field in which hands and surgical tools of both surgeons can be seen at the same time. \cite{rojas2019surgical} used a HoloLens mentoring system in which an expert surgeon can place virtual 3D annotations (surgical tools and incision guidance lines) which are seen by the student surgeon in real time  (Fig. \ref{fig:ponce2014emerging_fig3_and_rojas2019surgical_fig2} (b)). The authors reported improved information exchange between student and mentor, reduced number of focus shifts and reduced placement error. A similar mentoring is presented in \cite{rojas2020system}, where trainees performed leg fasciotomies and reported an improved surgical confidence. 

In contrast to telementoring, where the dialogue is continuous, teleconsultation (n = 5) focuses on a consultation based on-demand communication between colleagues. \cite{sauer2017mixed} explored potential benefits of using the HoloLens to establish a web-service based real-time video and audio communication with a remote colleague during visceral-surgical interventions (Fig. \ref{fig:sauer2017mixed_fig3}). In addition, the remote surgeon could mark anatomical structures within the surgical site using a tablet computer. \cite{borgmann2016feasibility} used a Google Glass for hands-free teleconsultation during different urological surgical procedures. Other examples use the Google Glass for consultation during reconstructive limb salvage \citep{armstrong2014heads} and orthopaedic procedures \citep{jalaliniya2017wearable}.

Applications where holographic 3D anatomy is displayed an intraoperative setup without trying to guide the surgical procedure are categorised as surgical anatomy assessment. This involves intraoperative assessment of preoperatively aquired patient anatomy that aids clinicial decision making without trying to guide the procedure itself.

\cite{sauer2017mixed} used a HoloLens based 3D visualisation of a liver cranio-ventral incl. tumor (Fig. \ref{fig:sauer2017mixed_fig3} (a)) to improve a surgeon's spatial understanding of the target anatomy during dissection of the liver parenchyma in complex visceral-surgical interventions. \cite{mitsuno2017intraoperative} used Moverio BT-200 smart glasses and registered holographic 3D face and facial bones surfaces (Fig. \ref{fig:pratt2018through_fig3_and_mitsuno2017intraoperative_fig8} (b)) to aid clinical decision making for more objective assessment of the improvement of a patient's body surface contour in plastic surgery. 

A further category of display shows preoperatively acquired 2D patient imaging data and medical records in the surgeon's field of view using AR rather than a separate monitor. \cite{borgmann2016feasibility} asked surgeons to rate their perceived usefulness of displaying patients' medical records and CT scans on a Google Glass during urological surgical procedures. They found that reviewing patient images was rated less useful, whereas reviewing medical records received a high rating. ~\cite{jalaliniya2015designing} used a Google Glass to view and manipulate X-ray and MRI images.

\begin{table*}
\centering
\begin{threeparttable}
\caption{Studies listed by OST-HMD, Surgical context and surgical procedure : Acronyms: SG: Surgical guidance. PS: Preoperative surgical planning. SAA: Intraoperative surgical anatomy assessment. ST: Surgical training. REV: Intraoperative review of preoperative 2D imaging and/or patient records. TELC: Teleconsultation during surgery. TELM: Telementoring. DOC: Intraoperative documentation. PM: Patient monitoring. Acronyms for surgical guidance applications: TP: Surgical tool placement. IO: Image overlay for navigation. SI: Screw insertion. NI: Needle insertion. CI: Catheter insertion. KWI: K-Wire insertion. EG: MIS Endoscopy guidance. SP: Stent-graft placement. DTG: Drill trajectory guidance. PN: Imaging probe navigation. SNN: Surgical saw navigation. CA: C-arm positioning guidance. RP: robot placement. DG: dissection guidance. AI: anatomy identification. }
\label{table:hmd_surgical_phase_surgical_applications}

\rowcolors{2}{white}{lightgray!20}
\begin{scriptsize}
\begin{tabular}{lp{0.11\linewidth}p{0.12\linewidth}p{0.5\linewidth}}
\toprule
    \rowcolor{white}
    \multicolumn{1}{c}{\textbf{Study}}
    & \multicolumn{1}{c}{\textbf{OST-HMD}}
    & \multicolumn{1}{c}{\textbf{Surgical context}}
    & \multicolumn{1}{c}{\textbf{Surgical procedure}} \\
  \midrule 
        \cite{armstrong2014heads} & Google Glass & TELC &  Reconstructive limb salvage procedures \\
         
        \cite{ponce2014emerging} & Google Glass & TELM & shoulder arthroplasty \\
        
        \cite{chen2015development} & nVisor ST60 &  SG (SI)  & Percutaneous implantation of sacroiliac joint screw \\
        
        \cite{katic2015system} & Custom Device & SG (DTG) & Dental implant surgery \\ 
        
        \cite{borgmann2016feasibility} & Google Glass & REV, ST, DOC, TELC & Different urological surgical procedures \\ 
        
        \cite{dickey2016augmented} & Google Glass & ST, TELM & Inflatable penile prosthesis placement \\
        
        \cite{liebert2016novel} & Google Glass & PM & Bronchoscopy \\
        
        \cite{wang2016precision} & nVisor ST60  & SG (SI) & Percutaneous implantation of sacroiliac joint screw \\ 
        
        \cite{stewart2016wearable} & Brother AirScouter WD-100G & SG (TP) & General intra-operative guidance (no concrete application, only measurement of attentiveness to the surgical field)\\
        
        \cite{kaneko2016ultrasound} & Moverio BT-200 &  SG (CI) & Central venous catheterisation under US guidance \\
        
       \cite{yoon2017technical} & Google Glass & SG (SI) & spine instrumentation (pedicle screw placement)\\
 
        \cite{jalaliniya2017wearable} & Google Glass & REV, TELC & Orthopaedic procedures \\
		
        \cite{li2017human} & HoloLens & PS, ST & Preoperative diagnosis \& planning of coronary heart disease\\
        
        \cite{kuhlemann2017towards} & HoloLens & SG (CI) & Interventional endovascular stenting of aortic aneurysm \\
		
        \cite{sauer2017mixed} & HoloLens & SAA, TELC & Visceral-surgical interventions \\
		
        \cite{mitsuno2017intraoperative} & Moverio BT-200 & SAA & Improvement of the body surface contour in plastic surgery.\\
        
        \cite{hiranaka2017augmented} & PicoLinker glasses & SG (KWI) & Fluoroscopy controlled K-wire insertion into femur \\
		 
        \cite{zou2017coronary} & Custom Device & PS & Preoperative diagnosis of coronary heart disease \\
        
        \cite{deib2018image} & HoloLens & SG (NI) & Percutaneous vertebroplasty, kyphoplasty and discectomy procedures \\
       
        \cite{Andres2018} & HoloLens & SG (KWI) & Percutaneous orthopaedic surgical procedures \\
        
        \cite{song2018endodontic} & HoloLens & SG (DTG) & Access cavity Preparation in Endodontic treatment \\
        
        \cite{condino2018build} & HoloLens & ST & Hip arthroplasty \\
        
        \cite{qian2018arssist} & HoloLens & SG (RP, TP, EG) & Increase the First Assistant's task performance during robot-assisted laparoscopic surgeries\\
        
        \cite{el-hariri2018augmented} & HoloLens & SG (TP) & Intra-operative bone localisation\\
        
        \cite{karmonik2018augmented} & HoloLens & PS & Identification of a hemodynamic scenario that predicts an aneurysm rupture\\
        
        \cite{lin2018holoneedle} & HoloLens & SG (NI) & Needle biopsy\\
        
        \cite{frantz2018augmenting} & HoloLens & SG (IO) & Neurosurgical applications \\
       
        \cite{pratt2018through} & HoloLens & SG (DG) & Vascular pedunculated flaps of the lower extremities (reconstruction surgery) \\
         
        \cite{unberath2018augmented} & HoloLens & SG (CA) & percutaneous orthopaedic procedures \\
        
        \cite{mahmood2018augmented} & HoloLens & ST & example: transesophageal echocardiography examination \\
        
        \cite{wu2018augmented} & HoloLens & SG (IO) & N/A \\
        
        \cite{boillat2019increasing} & Google Glass & DOC & Surgical time-out checklist execution \\
		
        \cite{meulstee2019toward} & HoloLens & SG (TP) & N/A \\ 
        \cite{gibby2019head} & HoloLens & SG (SI) & pedicle screw placement \\
        
        \cite{brun2019mixed} & HoloLens & PS & Repair for complex congenital heart disease \\
       
        \cite{deOliveira2019} & HoloLens & SG (IO) & Orthopaedic surgery (no specific procedure)\\
       
        \cite{fotouhi2019interactive} & HoloLens & SG (KWI) & C-arm fluoroscopy guided k-wire placement\\
        
        \cite{aaskov2019x} & HoloLens & SG (AI) & Identification of spinal anatomy underneath the skin \\
       
        \cite{guo2019online} & HoloLens & SG (IO) & General image-guided surgical navigation (no specific application)\\
       
        \cite{liebmann2019pedicle} & HoloLens & SG (SI) & Placement of pedicle screws in spinal fusion surgery\\
        
        \cite{liu2019augmented} & HoloLens & SG (CI) & transcatheter procedures for structural heart disease \\
       
        \cite{rojas2019surgical} & HoloLens & TELM & Abdominal incision \\
       
        \cite{rojas2020system} & HoloLens & TELM & Leg fasciotomy \\
        
        \cite{li2019mixed} & HoloLens & SG (TP) & liver tumor puncture \\
        
        \cite{pepe2019marker} & HoloLens & SG (IO) & Head and neck tumor resections \\
        
        \cite{zhou2019design} & HoloLens & SG (NI) & Seed implantation thoracoabdominal tumor brachytherapy  \\
        
        \cite{chien2019hololens} & HoloLens & SG (IO) & General SG (no specific surgical application)\\
         
        \cite{zhang2019preliminary} & HoloLens & SG (TP) & Craniotomy \\
         
        \cite{heinrich2019holoinjection} & HoloLens & SG (NI) & Needle-based spinal interventions \\
        
        \cite{wellens2019comparison} & HoloLens & PS & Nephron-sparing surgery \\
        
        \cite{fotouhi2019co} & HoloLens & SG (TP) & Percutaneous orthopaedic treatments \\
        
        \cite{rynio2019holographically} & Hololens & SG (SP) & Endovascular aortic repair \\
        
        \cite{zhou2019towards} & Magic Leap One & SG (PN) & Tooth decay management \\
         
        \cite{pietruski2019supporting} &  Moverio BT-200 & SG (SSN) &  Mandibular resection\\
        
        \cite{schlosser2019exploratory} & Vuzix M300 & PM & None \\
        
        \cite{fotouhi2020reflective} & HoloLens & RP & Set up of robotic arms by surgical staff (especially minimally invasive gastrectomy (abdominal surgery))\\
         
        \cite{pelanis2020use} & HoloLens & PS & Liver resection \\
         
        \cite{nguyen2020augmented} & HoloLens & SG (IO) & Neurosurgical applications \\ 
         
        \cite{zhou2020surgical} & HoloLens & SG (NI) & Seed implantation thoracoabdminal brachytherapy \\
         
        \cite{baum2020augmented} & HoloLens & ST & Neurosurgical burr hole localisation \\
        
        \cite{al2020effectiveness} & HoloLens & SG (EG) & Ureteroscopy \\
         
        \cite{pietruski2020supporting} & Moverio BT-200 & SG (SSN) & Free fibula flap \\ 
        
        \cite{liounakos2020head} & Moverio BT-300 & SG (EG) &  Percutaneous endoscopic lumbar discectomy  \\
        
        
        \cite{gnanasegaram2020evaluating} & HoloLens & ST & N/A \\
        
        \cite{sun2020high} & HoloLens & SG(CI) & External ventricular drainage (EVD) \\
        
        \cite{park2020three} & HoloLens & PS & Endovascular procedures \\
        
        \cite{mendes2020pinata} & Arzyon headset & ST & Central venous catheterisation \\
         
        \cite{laguna2020assessing} & HoloLens & PS & Repair of complex  paediatric elbow fractures \\ 
        
        \cite{dallas2020comparing} & HoloLens & PS & Complex surgical procedures \\
        
        \cite{zafar2020evaluation} & HoloLens & ST & No direct surgical procedure (teaching of dental anatomy)  \\
         
\multicolumn{4}{r}{(continued on next page)}
\end{tabular}
\end{scriptsize} 
\end{threeparttable}
\end{table*}
\begin{table*}
\ContinuedFloat
\centering
\begin{threeparttable}
\caption*{Table \thetable \hspace{0.2cm} (continued)}
\rowcolors{2}{white}{lightgray!20}
\centering
\begin{scriptsize}
\begin{tabular}{lp{0.11\linewidth}p{0.12\linewidth}p{0.5\linewidth}}
\toprule
    \rowcolor{white}
    \multicolumn{1}{c}{\textbf{Study}}
    & \multicolumn{1}{c}{\textbf{OST-HMD}}
    & \multicolumn{1}{c}{\textbf{Surgical context}}
    & \multicolumn{1}{c}{\textbf{Surgical procedure}} \\
  \midrule       
        \cite{fitski2020mri} & HoloLens & PS & Nephron-Sparing Surgery in Wilms' Tumor Surgery \\
        
        \cite{schoeb2020mixed} & HoloLens & ST & Urologic surgical procedures (bladder catheter placement) \\
        
        \cite{luzon2020value} & HoloLens & SG (DG) & Right colectomy with extended lymphadenectomy \\
        
        \cite{matsukawa2020smart} & PicoLinker glasses & SG (SI) & Single-segment posterior lumbar interbody fusion \\
        
        \cite{yang2020feasibility} & HoloLens & SG (NI) & Transjugular intrahepatic portosystemic shunt (TIPS) \\
        
        \cite{li2020smartphone} & HoloLens & SG (NI) & Percutaneous needle interventions \\
        
        \cite{kumar2020use} & HoloLens & PS & Example use cases: laparoscopic liver resection and congenital heart surgery \\
         
        \cite{li2020clinical} & HoloLens & SG (DG), PS, TELC, ST & Laparoscopic partial nephrectomy / Laparoscopic radical nephrectomy \\ 
        
        \cite{gibby2020use} & HoloLens & SG (NI) & Percutaneous image-guided spine procedures \\
        
        \cite{gu2020feasibility} & HoloLens & SG (DTG) &  Total shoulder arthroplasty \\
         
        \cite{galati2020experimental} & HoloLens & SAA & Open Abdomen Surgery \\
         
        \cite{viehofer2020augmented} & HoloLens & SG (SNN) & Hallux Valgus correction \\
        
        \cite{dennler2020augmented} & HoloLens & SG (SI) & Spinal instrumentation \\
        
        \cite{kriechling2020augmented} & HoloLens & SG (KWI) & Reverse total shoulder arthroplasty (RSA)  \\
        
        \cite{zorzal2020laparoscopy} & Metavision Meta 2 & SG (EG) & Laparoscopic procedures \\
         
        \cite{cartucho2020multimodal} & HoloLens & SAA & N/A \\ 
        
        \cite{rojas2020evaluation} & HoloLens & TELM & Cricothyroidotomy \\
        
        \cite{scherl2020augmented} & HoloLens & SG (IO) & Surgery of the parotid gland \\
         
        \cite{creighton2020early} & HoloLens & SG (IO) & Lateral Skull Base Surgery \\
        
        \cite{jiang2020hololens} & HoloLens & SG (DG) & Perforator flap transfer \\
        
        \cite{sun2020fast} & HoloLens & SG (IO) & Mandibular reconstruction \\
\bottomrule
\end{tabular}
\end{scriptsize}
\end{threeparttable}\end{table*}

\section{AR visualisations}
Conventional computer-assisted surgery uses different types of visualisations to aid preoperative planning or intraoperative procedures and a similar range of visualisations have been adopted for AR-assisted applications (see table \ref{table:appendix}).

\begin{figure}[ht!]
\resizebox{\columnwidth}{!}{
\begin{tikzpicture}
\begin{axis}[
symbolic x coords={, PM, IM, II, PI, IV, IND, COMM, PV, DOC, },
    ymin=0,
    ymax=69,
    ybar=-\pgfplotbarwidth,
    x=\pgfplotbarwidth,
    bar width=35pt,
	ymajorgrids,
	xmajorgrids,
	nodes near coords,
	ylabel={\#articles}
]
\addplot coordinates{(PM, 66)}; %
\addplot coordinates{(IM, 13)}; %
\addplot coordinates{(II, 12)}; %
\addplot coordinates{(PI, 11)}; %
\addplot coordinates{(IV, 5)}; %
\addplot coordinates{(IND, 3)}; %
\addplot[red,fill] coordinates{(COMM, 2)}; %
\addplot[orange,fill] coordinates{(PV, 2)}; %
\addplot[cyan,fill] coordinates{(DOC, 1)}; %
\legend{Preoperative Model, Intraoperative Model, Intraoperative Image, Preoperative Image, Intraoperative live streaming video, Intraoperative Numerical Data, 
2D plane with video communication, Preoperatively recorded video, Documents }
\end{axis}
\end{tikzpicture}
}
\caption{Distribution of included articles by \textbf{type of AR visualisation}}
\label{fig:distrib_AR_visualisations}
\end{figure}
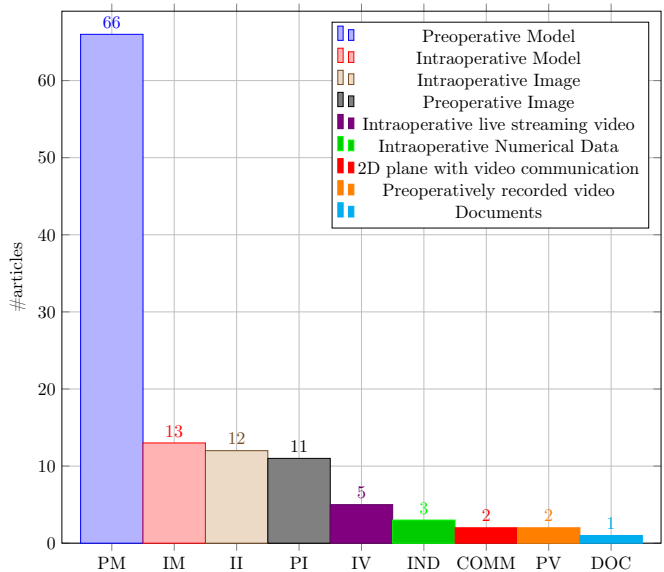

Fig. \ref{fig:distrib_AR_visualisations} shows the distribution of 
articles by type of AR visualisation. The majority of articles use preoperative models (n = 66), usually consisting of 3D reconstructed patient anatomy generated from CT or MRI imaging content, sometimes in conjunction with preoperative planning components. \cite{liebmann2019pedicle} used holographic preoperatively planned screw trajectories and drill entry points to aid pedicle screw placement in spinal fusion surgery. \cite{pratt2018through} investigated the usefulness of CT reconstructed 3D patient leg models including bony, vascular, skin and soft tissue structures, vascular perforators and a surrounding bounding box that facilitated manual registration. 

We also consider non-anatomical content as a preoperative model such as holographic user interaction menus or graphical annotations. \cite{rojas2020system}, for example, used graphical annotations of incision lines and a model of surgical tools in a telementoring system. \cite{condino2018build} implemented a virtual menu with toggle buttons for a  hybrid simulator for orthopaedic open surgery training.

Applications where 3D visualisations are generated intraoperatively in order to take updated live information into account, usually for surgical guidance, we refer to as intraoperative model visualisation (n = 13). \cite{katic2015system} used live drill trajectory guidance information such as position and depth of dental drill and injury avoidance warnings in dental implant surgery.
\cite{lin2018holoneedle} investigated utility aspects of intraoperatively generated needle visualisations such as needle position, orientation, shape and a tangential ray during needle biopsy.

Live intraoperative images (n = 12) can be displayed in a surgeon's field of view using AR in order to have crucial patient data available without the need to look at a separate monitor. \cite{deib2018image} displayed radiographic images to aid percutaneous vertebroplasty, kyphoplasty and discectomy procedures. \cite{qian2018arssist} used an endoscopy visualisation in the form of a 3D plane with video streaming content that aimed to increase the first assistant's task performance in robot-assisted laparoscopic surgery. \cite{fotouhi2019interactive} explored potential benefits of holgraphic C-arm interventional X-ray images registered to the C-arm view frustrum for guided k-wire placement in fracture care surgery.

The standard method of viewing preoperative images on a separate monitor away from the surgical site is often cited as a reason for pursuing AR guidance. Holographic visualisation of preoperative images (n = 5) was proposed to allow visualisation on or near the surgical site. \cite{song2018endodontic} incorporated 2D radiographic images with guidance information in their HoloLens-based endodontic treatment approach. \cite{rynio2019holographically} used 2D images with volume rendering, arterial diameters and planning notes to support endovascular aortic repair.

The remaining categories of AR visualisations we identified in this review have only a few applications. Intraoperative live video streaming (n = 5) is mostly used in telementoring applications. \cite{ponce2014emerging} used a hybrid image approach in which the mentee's surgical field is combined with the hands of the remote expert surgeon. \cite{dickey2016augmented} presented an application in which an interactive video display is visible to the mentee that shows a cursor moved by the supervising physician. Intraoperative numerical data (n = 3) is usually displayed as a 2D plane containing numerical data that aid clinical decision making or surgical guidance. \cite{pietruski2019supporting} displayed a cutting guide deviation coordinate system supporting a surgeon during mandibular resection. \cite{schlosser2019exploratory} implemented a patient monitoring application comprising a holographic 2D screen that shows patient heart rate, blood pressure, blood oxygen saturation and alarm notifications. Another AR visualisation category uses a 2D plane with video communication software (n = 2) and has been applied in reconstructive limb salvage procedures \citep{armstrong2014heads} and orthopaedic procedures \citep{jalaliniya2017wearable}. Preoperatively recorded video (n = 2) was explored by \cite{dickey2016augmented} as a video guide during surgical training. \cite{armstrong2014heads} used holographic visualisation of documents (n = 1), with articles from a senior author being displayed in the surgical field of view.

\section{Validation of AR}
\label{section:experiments}
All papers included in this review perform some kind of experiments to verify usability and the associated potential utility of their proposed OST-HMD assisted surgery solution. In this section we analyze the conducted experiments including a categorisation into an either quantitative or qualitative evaluation. An overview can be found in in the \textit{Experiments} column of table~\ref{table:appendix}.

\subsection{Experimental setting}

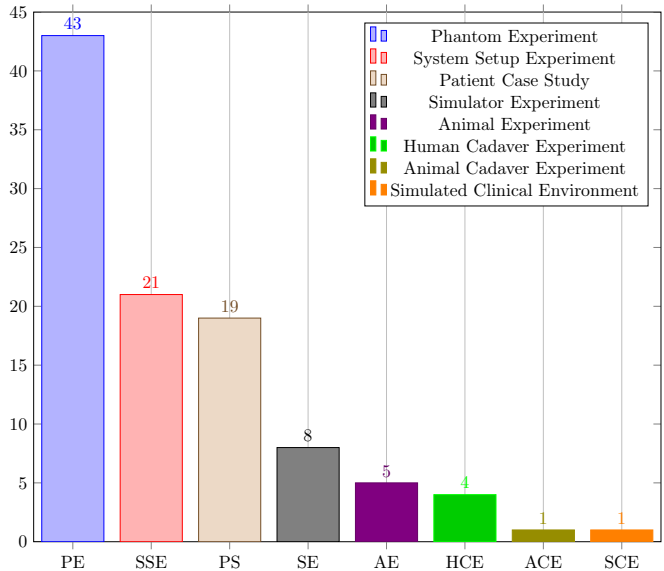
\begin{figure}[ht!]
    \centering
    \resizebox{\columnwidth}{!}{
        \begin{tikzpicture}
\begin{axis}[
    ybar=-\pgfplotbarwidth,
    bar width=35pt,
    xmin=-0.999 ,xmax=15,
    ymin=0, ymax=45,
    xtick=\empty,
    ybar=-\pgfplotbarwidth,
	nodes near coords,
    extra x ticks={0, 2, 4, 6, 8, 10, 12, 14 },
    extra x tick labels={PE, SSE, PS, SE, AE, HCE, ACE, SCE },
    extra x tick style={
           grid=major,
           tick label style={rotate=0} 
           }
    ]

\addplot coordinates {(0, 43)}; 
\addplot coordinates {(2, 21)}; 
\addplot coordinates {(4, 19)}; 
\addplot coordinates {(6, 8)}; 
\addplot coordinates {(8, 5)}; 
\addplot coordinates {(10, 4)}; 
\addplot[olive,fill] coordinates {(12, 1)}; 
\addplot[orange,fill] coordinates {(14, 1)}; 

\legend{Phantom Experiment, System Setup Experiment,  Patient Case Study, Simulator Experiment, 
Animal Experiment, Human Cadaver Experiment, Animal Cadaver Experiment, Simulated Clinical Environment}

\end{axis}
\end{tikzpicture}
    }
    \caption{Experimental setting, from phantom to animal to clinical studies. Phantom studies dominate and though a number of clinical case studies have been reported (19), we are some way from proving clinical effectiveness of OST-HMDs at present. }
    \label{fig:types_of_experiments}
\end{figure}

Phantom experiments dominate the list of papers (n = 43). Phantoms may be stylistic or try to mimic anatomical correct structures and are either self-made, 3D printed or acquired from specialised companies. Researchers can test their developed methods on phantoms without involving real human or animal anatomy. \cite{chen2015development} used a 3D-printed cranio-maxillofacial model to verify the registration accuracy of their presented surgical navigation system, and a 3D pelvis model to test their navigation system. \cite{deib2018image} incorporated a lumbar spine phantom into the validation of their presented application for image guided percutaneous spine procedures. A guidance approach for pedicle screw placement, developed by \cite{gibby2019head}, was tested using a phantom consisting of L1-L3 vertebrae in opaque silicone that mimics tissue properties.

System setup experiments (n = 21) don't use realistic target anatomy structures but verify the system's intrinsic characteristics by conducting accuracy experiments in specific areas, such as system registration and calibration. \cite{Andres2018}, for example, test the calibration step of their presented OST-assited fluoroscopic x-ray guidance system that uses a multimodal fiducial. The calibration experiment consists only of a HoloLens, a C-arm and a multimodality marker. \cite{fotouhi2019interactive} conducted a similar experiment incorporating a hand-eye calibration experiment including a HoloLens, a C-arm and an optical tracker in their system that provides spatially aware surgical data visualisation. In order to verify the calibration accuracy of their proposed online calibration method for the HoloLens, \cite{guo2019online} used 
a calibration box with visual markers, a tracking device based on computer vision.

Patient case studies (n = 19) present surgical procedures that were tested on one or more patients. \cite{yoon2017technical} validated their OST-assisted spine instrumentation approach in which neuronavigation images were streamed onto a Google Glass on 10 patients. \cite{mitsuno2017intraoperative} tested their intraoperative body surface improvement approach on 8 patients, each with a different diagnosis. These clinical evaluations are very useful, but further studies will be required to establish clinical effectiveness and to demonstrate improved patient outcome.

The remaining five types of experiments that have been identified in this review have a comparatively small number of associated articles. Simulator experiments (n = 8) take advantage of available simulation hardware allowing researchers or surgeons to mimic specific surgical procedures. \cite{mahmood2018augmented} used a physical simulator model (Fig.~\ref{fig:mahmood2018augmented_3_and_7}, section \ref{section:other_surgical_contexts}) that allows users wearing a HoloLens to simulate a transesophageal echocardiography (TEE) examination. 

Animal experiments (n = 5) involve living animals that are anaesthetised and enable surgeons to test surgical applications under realistic conditions that consider physiological aspects such as respiratory motion. \cite{zhou2019design} and \cite{zhou2020surgical}  tested their surgical navigation system for LDR brachytherapy on a live porcine model (Fig.~\ref{fig:liebmann2019pedicle_fig5_b__d_and_zhou2019design_fig1} (c), section \ref{section:surgical_context}). \cite{li2019mixed} performed a similar in vivo test of their respiratory liver tumor puncture navigation system that takes respiratory liver motion into account. An animal cadaver experiment (n = 1) was also performed by \cite{katic2015system}, who used a pig cadaver to test their application for intraoperative guidance in dental implant surgery.

Human cadaver experiments (n = 4) have the inherent benefit of allowing surgeons to test novel surgical procedures on real anatomic structures without risk to patients. \cite{wang2016precision}, for example, used six frozen cadavers with intact pelvises to investigate a novel method for insertion of percutaneous sacroiliac screws.

Finally, \cite{jalaliniya2017wearable} proposed a simulated clinical environment (n = 1) to test clinical infrastructure elements and workflows rather than surgical procedures. A Google Glass based wearable personal assistant that allows surgeons to use a videconferencing application,  visualise patient records and enables touchless interaction with preoperative X-ray and MRI images displayed on a separate screen without the need to use mouse or keyboard. The application was tested in different clinical setups comprising a simulation doll, human actors and real surgeons and nurses.

\subsection{Evaluation methods}
Evaluations may be quantitative experiments that collect measurable data such as registration accuracy or qualitative experiments gather descriptive information such as surgeons' observations or opinions that cannot be measured. Most of the articles in this review contain some sort of quantitative experiments (n = 88), whereas qualitative experiments have much fewer associated articles (n = 11). 

Quantitative experiments include registration accuracy evaluation  \citep{chen2015development, condino2018build, gibby2019head}), calibration accuracy evaluation \citep{Andres2018, fotouhi2019interactive, qian2018arssist}) or intraoperative guidance verification such as tool positioning \citep{stewart2016wearable} or guide wire placement \citep{liebmann2019pedicle}. Experiments in which a user has to give specific survey-based feedback is also classed as quantitative where the survey is predetermined and can be evaluated on a numerical basis. 

Qualitative experiments are usually questionnaire based in which the participants detail specific observations that cannot be evaluated numerically. \cite{deib2018image}, for example, designed an experiment in which the user had to complete a questionnaire following a surgical image-guided spine procedure describing benefits, limitations and personal preferences.

\section{Registration and tracking in surgical AR}
\label{section:accuracy}
Whenever holographic anatomy visualisations need to be superimposed on respective patient anatomy, the question of registration accuracy arises. 
Registration refers to the establishment of a spatial alignment between the coordinate system of the patient space and the digital image space \citep{liu2017new}. In the context of AR-guided surgery it can be defined as achieving correspondence between superimposed visualisation and patient anatomy. Devices such as the HoloLens define their own coordinate system for the room and the user's head is tracked within this space. The registration process places the preoperative model in HoloLens coordinates. If an external tracking system is used a further alignment between the devices is required. Tracking and registration each have potential errors and should be considered separately. 

In most cases a rigid coordinate system transformation involving translation and rotation is optimised given some corresponding features \citep{wyawahare2009image}. The required accuracy of the established registration depends on the application. For OST-HMD AR-guided procedures deviations between visualisation and true target anatomy may lead to surgical errors resulting from misinterpreted spatial relationships. For OST-HMD AR, overall accuracy also depends on the user's perceptual accuracy. 

Appendix table \ref{table:appendix} lists the main reported accuracy results of all included articles and the associated type of conducted experiment or experiments. Because of wide variation in experimental setup and different accuracy metrics used in the literature, direct comparison of articles based on the reported accuracy is difficult. Some articles report specific registration accuracy experiments \citep{chen2015development, gibby2019head, li2019mixed, nguyen2020augmented, heinrich2019holoinjection}, while others report the accuracy of specific experimental guidance task results that result from a preceding registration \citep{wang2016precision, stewart2016wearable, lin2018holoneedle, hiranaka2017augmented}. 

A number of papers consider manual alignment of the virtual model by the surgeon for registration. When matching corresponding features the most common methods are point-based landmark registration and surface registration \cite{liu2017new}.

\subsection{Manual alignment}
\label{sec:manual_alignment}
\cite{pratt2018through} propose manual registration for extremity reconstruction using the HoloLens. Manual registration directly aligns the model to the HoloLens coordinate system, so no further tracking calculation is required. Also, since the alignment is achieved by the user's and to their satisfaction, no correction for individual's 3D perception is needed. \cite{fotouhi2020reflective}
propose manual registration for virtual-to-real alignment of a robotic arm that uses two reflective AR displays (Fig. \ref{fig:qian2018arssist_fig4f_and_fotouhi2020reflective_fig4} (b)). The reflective AR displays act as holographic mirrors that allow the first assistant to see the virtual robot arm from multiple perspectives and therefore act as a registration aid. Experiments showed that using the reflective AR displays improved the accuracy from $30.2 \pm 23.9$ mm to $ 16.5  \pm  11.0 $ mm.
\cite{nguyen2020augmented} compared three manual registration methods for neuronavigation using the HoloLens: tap to place, 3-point correspondence matching and keyboard control. The authors also presented a novel statistics based method allowing researchers to quantify registration accuracy for AR-assisted neuronavigation approaches. The keyboard method was found to be the most accurate (for detailed accuracy results see appendix table \ref{table:appendix}).

\begin{figure*}[ht!]
\centering
\resizebox{\textwidth}{!}{
\begin{tabular}{cc}
 \includegraphics[height=0.5\textwidth]{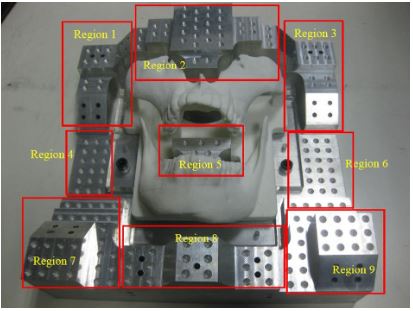} &
 \includegraphics[height=0.5\textwidth]{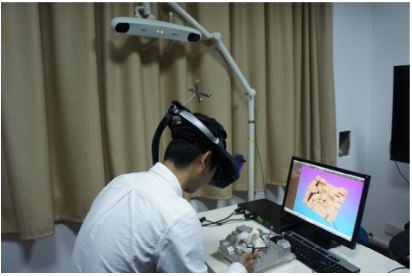} \\
(a) & (b) \\
\multicolumn{2}{c}{
\includegraphics[height=0.5\textwidth]{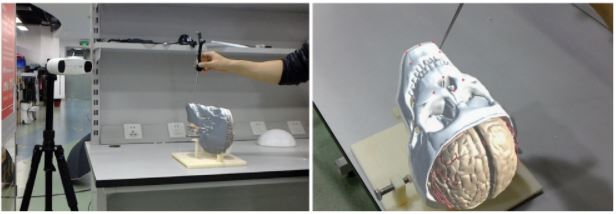}} \\
\multicolumn{2}{c}{
(c)}
\end{tabular}
}
\caption{Accuracy verification experiment examples using optical trackers: (a) Accuracy verification block including a metal base with taper holes (for distance and angular error measuring) and 3D-printed cranio-maxillofacial model. (b) A user is conducting the accuracy verification experiment using the accuracy verification block, a tracked calibration tool and tracked OST-HMD \textit{(source: \cite{chen2015development}Fig. 7 and 8(c))}. Registration accuracy validation using a 3D-printed skull with 10 landmarks (red dots) and a k-wire with attached optical marker \textit{(source: \cite{li2019mixed} Fig. 6)}}
\label{fig:chen2015development_fig7_and_8_and_li2019mixed_fig6}
\end{figure*}

\begin{figure*}[ht!]
\centering
\resizebox{\textwidth}{!}{
\begin{tabular}{ccc}
 \includegraphics[height=0.3\textwidth]{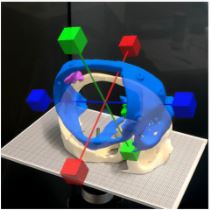} &  
 \includegraphics[height=0.3\textwidth]{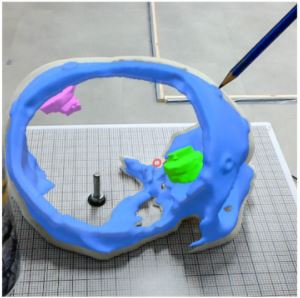} & 
 \includegraphics[height=0.3\textwidth]{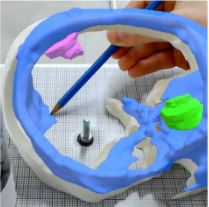} \\
(a) & (b) & (c)\\
\end{tabular}
}
\caption{Registration accuracy verification using a sheet of millimeter paper: (a) Manual and point-based registration: Virtual axes allow the user to translate and rotate a human skull model in order to align it with a phantom. Fiducial markers serving as registration aids are present on both the virtual model and the phantom. (b) Localisation accuracy measurement is realised by placing the tip of a stylus into the center of a holographic fiducial marker. (c) By calculating the difference in similar points the perceived hologram drift is measured.  \textit{(source: \cite{frantz2018augmenting} Fig. 4a, 4b and 5a).}}
\label{fig:frantz2018augmenting_fig4a_4b_5a}
\end{figure*}

\cite{frantz2018augmenting} presented a neuronavigation approach which is based on manual registration using fiducial markers. Users can manually register a holographic visualisation of a 3D reconstructed CT scan human skull model to its physical counterpart via the help of virtual axes (Fig. \ref{fig:frantz2018augmenting_fig4a_4b_5a} (a)). Registration accuracy was measured by both localisation accuracy (Fig. \ref{fig:frantz2018augmenting_fig4a_4b_5a} (b)) and perceived holographic drift (Fig. \ref{fig:frantz2018augmenting_fig4a_4b_5a} (c)). The mean perceived holographic drift of the manual registration was $4.39$ $\pm$ $1.29$ mm. Maintaining hologram registration via continuous tracking of a marker resulted in a lower perceived hologram drift of $1.4$ $\pm$ $0.67$ mm.

These manual methods may not be of sufficient accuracy to meet the clinical requirements for guidance of surgical dissection, but the ability to orient structures can improve spatial awareness and may be useful in broader surgical decision making.

\subsection{Point-based registration}

Point-based registration matches corresponding pairs of fiducial points from one coordinate system to another. External fiducial markers may be attached to specific patient anatomy, such as bony structures in orthopaedic surgery or the skull in neurosurgery. Alternatively existing anatomical landmarks may be used. The same virtual fiducial points are usually marked using an external tracking device and can also be displayed on the holographic 3D anatomy model. A common accuracy measure for point-based methods is the fiducial registration error (FRE), which is the residual error of the mismatch between pairs of corresponding points after alignment. A better metric with more clinical relevance is the target registration error (TRE) at the surgical target \citep{seginer2011rigid}.

\cite{chen2015development} perform point-based registration as an initial alignment before surface-based refinement (section \ref{sec:surface_based_registration}) and conducted an accuracy experiment using a verification block (Fig. \ref{fig:chen2015development_fig7_and_8_and_li2019mixed_fig6} (a)) using an optical tracking system with reflective markers \ref{fig:chen2015development_fig7_and_8_and_li2019mixed_fig6} (b)). The authors reported mean distance and angular errors of $0.809 \pm 0.05$ $mm$ and $1.038^{\circ}\pm0.05^{\circ}$ respectively. 
Addressing the problem of incorrect needle placement and associated failed tumor ablation, \cite{li2019mixed} proposed a manual registration method using a HoloLens with an optical tracker to superimpose 3D liver models on patients for liver tumor puncture navigation. Optical markers rigidly attached to the HoloLens, anatomical marks on the patient and a k-wire with attached reflective spheres serving as an optical marker are used for an initial manual registration step. The tracked k-wire is then used for automatic temporal registration during the procedure. The authors performed a registration accuracy validation experiment using a 3D-printed skull with 10 landmarks (Fig. \ref{fig:chen2015development_fig7_and_8_and_li2019mixed_fig6} (c)) and reported an average target registration error of $2.24$ mm.

\begin{figure*}[ht!]
\centering
\resizebox{\textwidth}{!}{
\begin{tabular}{cc}
 \includegraphics[height=0.45\textwidth]{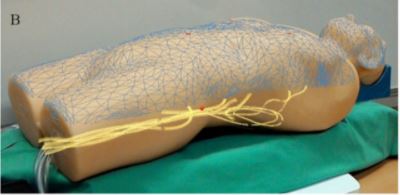} &  
 \includegraphics[height=0.45\textwidth]{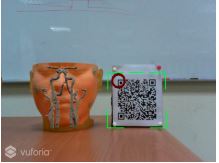} \\
(a) & (b) \\
\end{tabular}
}
\caption{(a) Point-based registration: Human body surface mesh including vessel tree registered to a phantom by landmarks, where surface registration to the HoloLens surface failed \textit{(source: \cite{kuhlemann2017towards} Fig. 1)}. (b) Surface registration result: Dummy Head with superimposed 3D CT scan reconstruction of head and intracranial vasculature. HoloLens camera detection of the QR code provides tracking \textit{(source: \cite{wu2018augmented}, part of Fig. 9b)}}
\label{fig:kuhlemann2017towards_fig1_and_wu2018augmented_fig_9b}
\end{figure*}

Another point-based registration approach for catheter navigation was presented by \cite{kuhlemann2017towards} and tested on a human body phantom (Fig. \ref{fig:kuhlemann2017towards_fig1_and_wu2018augmented_fig_9b} (a)): A CT reconstructed 3D body surface mesh including marching cubes segmentation of a vessel tree was registered to a body phantom using landmarks with a reported accuracy of $4.34\pm 0.709$ mm (FRE).

\subsection{Surface registration}
\label{sec:surface_based_registration}
Point-based registration is an alignment process that matches anatomical or fiducial landmarks. Surface registration offers the possibility of alignment without specific fiducial markers. Using a laser range scanner or a tracked probe, a point cloud is collected from the surface of the patient's target anatomy (e.g. the head) \citep{liu2017new}. Another surface or point cloud is derived from the image space and an algorithm is then used to match both point clouds. Most surface registration methods require a coarse manual or point-based registration step to place the image-based point cloud must be placed close to the target registration pose before the algorithm proceeds. Iterative closest point (ICP) is a popular realisation of a surface based registration and has been applied in several of our selected articles. 

The HoloLens internal tracking method produces a generated surface mesh and \cite{kuhlemann2017towards} investigated whether this could be used for surface registration. A CT scan derived body surface was matched to the HoloLens surface mesh. But the HoloLens mesh resolution was found to be too coarse. In addition, \cite{frantz2018augmenting} also reported that the HoloLens' built-in spatial mesh and simultaneous localisation and mapping (SLAM) system is unsuitable for registration and subsequent tracking due to the low vertex density and surface bias of the generated mesh and uncertainty in the SLAM realisation. \cite{wu2018augmented} presented an improved version of the ICP algorithm for medical image alignment that aims to provide a global optimum via a stochastic perturbation. A dummy head alignment test revealed an average target registration error of $< 3$ mm. Fig. \ref{fig:kuhlemann2017towards_fig1_and_wu2018augmented_fig_9b} (b) shows an example registration result.

\subsection{Other registration methods}
\label{section:other_registrations}
Other types of registration have also been explored. \cite{liu2019augmented} applied a Fourier transformation based registration method in their intraoperative guidance approach for structural heart disease for transcatheter procedures. The authors used a 3D reconstructed spine image and a segmented spine from an intraoperative fluoroscopy to calculate a Fourier-based scale and rotational shift which was then used to register the fluoroscopic image to the respective 3D model of the spine. The Fourier based registration achieved an accuracy of $0.42\pm 0.02$mm.

A HoloLens specific marker-less automatic registration method for maxillofacial surgery is presented by \cite{pepe2019marker}. Their algorithm accesses the HoloLens' built-in RGB camera and extracts facial landmarks from the camera's video stream. Via known virtual-to-real world transformations of the landmarks and spatial mapping information from the HoloLens' Spatial Mapping API, the algorithm then computes the registration. The achieved average positioning error of the x, y, z axes was $ 3.3 \pm 2.3$ mm \,  y: -$4.5 \pm 2.9$ mm \ and z: -$9.3 \pm 6.1$ mm respectively. 

\begin{table}
\centering
\resizebox{\columnwidth}{!}{%
\begin{threeparttable}
\caption{Papers by tracking method} 
\label{table:papers_by_tracking_method}
\rowcolors{2}{white}{lightgray!20}
\begin{scriptsize}
\begin{tabular}{llll}
\toprule
\rowcolor{gray!30}
\textbf{Tracking method totals} & & \textbf{Tracking marker totals} &\\
\midrule
& & & \\
\textbf{External tracker} & 19 & & \\
NDI Polaris & 11 & Reflective spheres & 14 \\
NDI EM/Aurora &	4 &	EM	& 4 \\
OptiTrack &	1 &	& \\
PST Base &	1 & & \\
VICON &	1  & & \\
Custom webcam tracker &	1 & Coloured catheter segments & 1	\\
 & & & \\
\textbf{Tracking with HMD camera} &	20 &	\textbf{Optical Markers} &	18 \\
HoloLens &	17 &	AprilTag &	1 \\
Other &	3 &	Aruco &	1 \\
 & & ARToolkit &	3 \\
& &	Custom	& 4 \\
& &	Vuforia	& 9 \\
& & & \\
\textbf{Markerless tracking}	& 11 & & \\

\bottomrule
\end{tabular}
\end{scriptsize} 
\end{threeparttable}
}
\end{table}

\subsection{Tracking}
Having established a registration, any subsequent motion of either the patient or the surgeon must be tracked to maintain the alignment. A summary of tracking methods is given in table~\ref{table:papers_by_tracking_method}.

\subsubsection{Markerless tracking}
The HoloLens inherently tracks the surgeon's head and providing the patient position is fixed within the operating room, this may be a sufficient method in itself. Eleven papers use only the HoloLens tracking and these are associated with the manual registration process described in section~\ref{sec:manual_alignment}. The advantage of this method is that no external measurement device is required and no markers need to be physically attached to the patient, hence the name markerless tracking. This can be a significant advantage in terms of sterility, convenience and operative workflow integration. However, the accuracy is user dependent and may not be sufficient for some surgical tasks.
\cite{pratt2018through} and \cite{scherl2020augmented} use manual alignment to the anatomy, whereas \cite{creighton2020early} register to fiducial markers for guidance of targets in the skull base.

\subsubsection{Tracking of markers using the OST-HMD device}
OST-HMD devices such as the HoloLens incorporate cameras into their tracking process. These cameras can be used to track surface features or markers placed in the surgical field, accounting for 20 of our papers. It is common for these markers to be small planar identifiable markers modelled on QR codes. Several quite similar free libraries are available for this purpose, including Aruco, ARToolkit and AprilTags. \cite{Andres2018} use ARToolkit markers that are also visible in X-ray to align to fluoroscopic views for orthopaedics. \cite{liebmann2019pedicle} use the stereo HoloLens camera sensors in research mode to track planar sterile markers for pedicle screw navigation. Some authors use their own custom markers, such as the cube and hexagonal markers used by \cite{zhou2019design} in their system for brachytherapy. The commercial Vuforia package can also be used to track any planar printed image and accounts for half of the marker-based tracking through the OST-HMD (9 papers).

One advantage is that the OST-HMD camera's position is relative to the surgeon, so no extra registration is needed and the direction of the camera shoud be towards the surgical field. While this can be effective, the resolution and field of view of the cameras may not  be best designed for tracking within the surgical target area.

\subsubsection{External tracking devices}
There are several commercially available devices that are able to track markers within the operating room. It is clear from table~\ref{table:papers_by_tracking_method} that Northern Digital Inc. (NDI) dominate this field, with the Polaris optical tracker accounting for 11 papers and their electromagnetic tracker, Aurora, a further four papers. \cite{li2019mixed} use the Polaris for liver biopsy in the presence of breathing, whereas \cite{kuhlemann2017towards} use EM tracking for endovascular interventions. Other system are optical and account for one paper each (OptiTrack, PST Base and VICON). Apart from one custom tracker based on a webcam for catheter tracking~\citep{sun2020high} all optical systems use passive reflective spherical markers.

It may be invasive to attach such markers rigidly to the patient, but such methods form part of several commercial image guidance systems and this is probably the most accurate way to achieve and maintain alignment.

\section{Human factors}
OST-HMDs are wearable technological devices that enable the user to visualise and/or interact with 3D virtual objects placed within their normal view of the world. These unfamiliar devices present a novel form of human-computer interaction (HCI) and their acceptability by surgeons will depend on HCI factors. 
Technological aspects, such as the size of the augmented field of view or system lag during streaming of video content, can affect user acceptance. But beyond these are human factors that may vary from user to user but are crucial to the utility of a technological interaction device. They encompass perceptual, cognitive and sensory-motor aspects of human behavior that drive the design of HCI interfaces to optimise operator performance \cite{papantoniou2016glossary}.

However, attempts to identify consistent generic human factors that capture basic human behavior and cognition that apply to the design of optical HCI systems has been problematic and HCI design guidelines incorporating consistent human factors have not yet been established. When addressing the negative side effects of HCI aspects only, human factors are sometimes considered as human limitations. Highlighting the aspect of human error, \cite{lowndes2014overview} addressed aspects of human factors and ergonomics in the operating room in general with a focus on MIS and found that most medical errors are a result of suboptimal system design causing predicable human mistakes.  They also state that despite efforts made by human factors and ergonomics professionals to improve safety in the operating room for over a century, increasingly complex surgical procedures and advances in technology mean that consideration of human interaction will be required to help users cope with increasing information content. We believe that similar safety aspects of human factors also apply in OST-HMD assisted surgical applications.

\subsection{Human factors in AR}
\label{section:human_factors_in_AR}
In more general non-surgical AR applications, human factors have played an important role and have been explored in the context of HCI. 
\cite{Livingston2005} evaluated human factors in AR in 2005 and found that apart from technological limitations, human factors are a major hurdle when it comes to translation of AR applications from laboratory prototypes into commercial products. To determine the effectiveness of AR systems requires usability verification, which led them to the following two research questions: \emph{1.) How to determine the AR user's key perceptual needs and the best methods of meeting them via an AR interface? 2.) Which cognitive tasks can be solved better with AR methods than with conventional methods?} 
They attempt to a solution for these two questions by conducting limited but well-designed tests aiming to provide insights into HCI-design aspects that lead to utility for perceptive and cognitive tasks. These consist of low-level perceptual tests of specific designed visualisations on the one hand and task-based tests that focus only on the well-designed part of the user interface.
In \cite{huang2012human}, Livingston points out that designing cognitive tasks for usability evaluation seems to be easier than designing low-level perceptual tasks, since cognitive tasks naturally arise from the given AR application, whereas it is rather challenging to design general low-level perceptual tasks that have wider applicability. He also states that the design of a perceptual task determines how generalisable the evaluation results are beyond the specific experimental scenario and indicates that a solution may be to design general perceptual tasks that verify the usability of hardware. Finding general perceptual tasks is not always easy when hardware limitations interfere with the task design. If the effect of a hardware related feature influences a user's cognition on top of his/her perception, the dependence on the perceptual task will be increased as well. An example for such an effect is system latency in a tracking device.

\subsection{HCI design considerations in OST-assisted surgery}
Though human factors in the context of HCI design considerations in OST-HMD assisted surgery are likely to be very important to the success of any system, only a few examples of such research can be found in the literature. In addition to the need for careful experimental design that allows a generalised result, there are also technical aspects that can be addressed to minimise unwanted human behavior when using OST-HMDs. Such technical aspects were explored by \cite{tang2003evaluation}, who evaluated human factors in variants of the \emph{Single point active alignment method (SPAAM)} for OST-HMD calibration that require human-computer interaction.
They aimed to answer the question why calibration of OST-HMDs is challenging for users; and found that human factors have a major impact on calibration error and therefore lead to significantly different accuracy results for different users. They proposed the following guidelines for the design of OST-HMD calibration procedures: \emph{1.) Calibration should not rely on head movements only, 2.) The user's head should be kept stabilised by minimizing extrinsic body movements} and \emph{3.) Careful consideration of the data collection sequence for the left and right eye so that calibration error does not bias towards a dominant eye.}

\cite{guo2019online} also described the importance of human factors in the context of OST-HMD calibration. They proposed an online calibration method for the HoloLens and concluded that 
the accuracy of their calibration method is difficult to measure objectively since
human factors impact the overall HCI experience as well as influencing the calibration accuracy.

The relationship between conscious and unconscious cognitive processes should be considered as well when considering the importance of human factors in HCI. \cite{jalaliniya2015designing} addressed the necessity to consider an egocentric interaction when designing wearable HCI systems and replaced the terms \emph{input} and \emph{output} with \emph{action} and \emph{perception}. According to the authors, an improved understanding of a human's perception, cognition and actions are necessary prerequisite when it comes to the design of a HCI system that offers better cognitive support. 

\subsection{Human factors identification}
We identified numerous human factors in the 91 included articles. Even though the majority of articles didn't use the term  \textit{human factors} explicitly, we included all user-related aspects described by the authors that have a potential impact on the acceptance, utility and performance of surgery with or without the proposed OST-HMD solution. 

We identified 34 human factors that are described in table \ref{table:human_factors_list}. The human factors are grouped into two categories: 1.) Human factors of conventional surgery that are addressed by OST-HMD AR and 2.) Persistent human factors that remain or are an inherent part of the proposed OST-HMD solutions. We further categorize them into three phases of user interaction that are defined by the US Food and Drug Administration as part of a medical device user interface in an operational context \citep{fda2016}: 1.) Information Perception (IP), where the information from the device is received by the user 2.) Cognitive Processing (CP), where the information is understood and interpreted and 3.) Control Actions (CA), where this interpretation leads to actions.

\begin{table*}
\centering
\begin{threeparttable}
\caption{Identified Human Factors, grouped into the categories 1.) Information Perception, 2.) Cognitive Processing and 3.) Control Actions} 
\label{table:human_factors_list}
\rowcolors{2}{white}{lightgray!20}
\begin{scriptsize}
\begin{tabular}{ll}
\toprule
\rowcolor{gray!30}
\textbf{Abbreviation} & \textbf{Human Factor}\\
\midrule
\multicolumn{2}{c}{\textbf{Information Perception}}\\
SPATIAL\_PERC & Spatial perception/awareness \\
INC & Inconvenience \\
DPPC & Missing/impaired depth perception  \\
EYE &  Individually different visual processing capabilities between dominant and non-dominant eye \\
COMF & Perceived comfort level when wearing OST-HMD \\
PER\_REAL\_AUG & Perception of spatial relationships between real and virtual objects \\
IMMR &  Personal degree of perceived immersion \\
\hline
\multicolumn{2}{c}{\textbf{Cognitive Processing}}\\
ATTN\_SHIFT & Attention switch between surgical site and separate computer monitor \\ 
MM & Error-prone and cognitively demanding mental mapping of 2D image data to 3D word \\
SLC & Steep learning curve \\
EXP\_OUTCOME & Influence of clinician's experience on surgical outcome \\
DIST & Distraction \\
INTPN\_2D\_DETAIL & Risk of incorrect interpretation of 2D image details  \\
INTRA\_OP\_NAV &  Impaired intraoperative navigation abilities due to absence of visual aids \\
COMM\_3D &  Personal 3D anatomical imagination capabilities affect communication between experts \\
CONF & Confidence \\
FRUS & Frustration  \\
SUBJ\_MEAS\_OUTCOME & Subjective measurement of surgical outcome  \\
EASE\_HCI &  Perceived degree of ease and intuitiveness of HCI \\
CLIN\_EXP\_2D &  Dependence on clinical experience for interpretation of 2D image data  \\
EMP\_EST\_2D &  Inaccurate empirical estimation of target locations in 2D anatomy images \\
ANAT\_PLN & Impaired anatomical understanding during preoperative planning due to 2D imaging data \\
CONC\_LS & Loss of concentration \\
MIP & Limited mental information processing abilities \\
STRESS & Experience of stress \\
ENG\_MOT & Engagement and motivation \\
PREF\_HOL &  Preferred degree of superimposition of 3D objects onto the surgical field (precise vs shifted superimposition) \\
USEF & Perceived usefulness of OST-HMD  \\
ANX &  Anxiety \\
\hline
\multicolumn{2}{c}{\textbf{Control Actions}}\\
VIS\_OPT & Selection of preferred mode of visualisation \\
SURG & Increased risk of surgical error  \\
HEC & Unfamiliar/cognitively demanding hand-eye coordination \\
TOOL\_ADJUST & Error-prone manual tool adjustment \\
FAT &  Visual Fatigue \\
\bottomrule
\end{tabular}
\end{scriptsize} 
\end{threeparttable}
\end{table*}

\begin{figure}[ht!]
\begin{tikzpicture}[scale=0.56]
\begin{axis}[
    width=16.5cm,
    height=7cm,
    nodes near coords,
    xmin=-0.999 ,xmax=55,
    ymin=0, ymax=45,
    xtick=\empty,
    ybar=-\pgfplotbarwidth,
    extra x ticks={0, 2, 4, 6, 8, 10, 12, 14, 16, 18, 20, 22, 24, 26, 28, 30, 32, 34, 36, 38, 40, 42, 44, 46, 48, 50, 52, 54, 56, 58, 60, 62, 64 },
    extra x tick labels={\textcolor{atomictangerine}{SPATIAL\_PERC (IP)}, \textcolor{atomictangerine}{INC (IP)}, \textcolor{atomictangerine}{DPPC (IP)}, \textcolor{atomictangerine}{EYE (IP)}, \textcolor{atomictangerine}{COMF (IP)}, \textcolor{amethyst}{ATTN\_SHIFT (CP)}, \textcolor{amethyst}{MM (CP)}, \textcolor{amethyst}{SLC (CP)}, \textcolor{amethyst}{EXP\_OUTCOME (CP)}, \textcolor{amethyst}{DIST (CP)}, \textcolor{amethyst}{INTPN\_2D\_DETAIL (CP)}, \textcolor{amethyst}{INTRA\_OP\_NAV (CP)}, \textcolor{amethyst}{COMM\_3D (CP)}, \textcolor{amethyst}{CONF (CP)}, \textcolor{amethyst}{FRUS (CP)}, \textcolor{amethyst}{SUBJ\_MEAS\_OUTCOME (CP)}, \textcolor{amethyst}{EASE\_HCI (CP)}, \textcolor{amethyst}{CLIN\_EXP\_2D (CP)}, \textcolor{amethyst}{EMP\_EST\_2D (CP)}, \textcolor{amethyst}{ANAT\_PLN (CP)}, \textcolor{amethyst}{CONC\_LS (CP)}, \textcolor{amethyst}{MIP (CP)}, \textcolor{amethyst}{STRESS (CP)}, \textcolor{amethyst}{ENG\_MOT (CP)}, \textcolor{airforceblue}{SURG (CA)}, \textcolor{airforceblue}{HEC (CA)}, \textcolor{airforceblue}{TOOL\_ADJUST (CA)}, \textcolor{airforceblue}{FAT (CA)}}, 
    extra x tick style={
           tick label style={rotate=90} 
           },
    ymajorgrids,
	xmajorgrids,
    ]
\addplot[atomictangerine, fill] coordinates{(0, 20)}; 
\addplot[atomictangerine!70, fill] coordinates{(2, 1)}; 
\addplot[atomictangerine!70, fill] coordinates{(4, 1)}; 
\addplot[atomictangerine!70, fill] coordinates{(6, 1)}; 
\addplot[atomictangerine!70, fill] coordinates{(8, 1)}; 

\addplot[amethyst!100, fill] coordinates{(10, 41)}; 
\addplot[amethyst!90, fill] coordinates{(12, 24)}; 
\addplot[amethyst!80, fill] coordinates{(14, 20)}; 
\addplot[amethyst!70,fill] coordinates{(16, 7)}; 
\addplot[amethyst!60,fill] coordinates{(18, 5)}; 
\addplot[amethyst!50,fill] coordinates{(20, 4)}; 
\addplot[amethyst!45,fill] coordinates{(22, 3)}; 
\addplot[amethyst!45,fill] coordinates{(24, 3)}; 
\addplot[amethyst!40,fill] coordinates{(26, 2)}; 
\addplot[amethyst!40,fill] coordinates{(28, 2)}; 
\addplot[amethyst!40,fill] coordinates{(30, 2)}; 
\addplot[amethyst!40,fill] coordinates{(32, 2)}; 
\addplot[amethyst!40,fill] coordinates{(34, 2)}; 
\addplot[amethyst!40,fill] coordinates{(36, 2)}; 
\addplot[amethyst!37,fill] coordinates{(38, 1)}; 
\addplot[amethyst!37,fill] coordinates{(40, 1)}; 
\addplot[amethyst!37,fill] coordinates{(42, 1)}; 
\addplot[amethyst!37,fill] coordinates{(44, 1)}; 
\addplot[amethyst!37,fill] coordinates{(46, 1)}; 

\addplot[airforceblue,fill] coordinates{(48, 23)}; 
\addplot[airforceblue!80,fill] coordinates{(50, 18)}; 
\addplot[airforceblue!60,fill] coordinates{(52, 2)}; 
\addplot[airforceblue!60,fill] coordinates{(54, 2)}; 
\end{axis}

\end{tikzpicture}
\caption{Distribution of human factors of conventional non-AR surgical approaches alleviated by the use of OST-HMD AR, grouped into the three categories \textcolor{atomictangerine}{1. Information Perception (IP)}, \textcolor{amethyst}{2. Cognitive Processing (CP)}, \textcolor{airforceblue}{3. Control Actions (CA)}}
\label{fig:distribution_addressed_human_factors}
\end{figure}
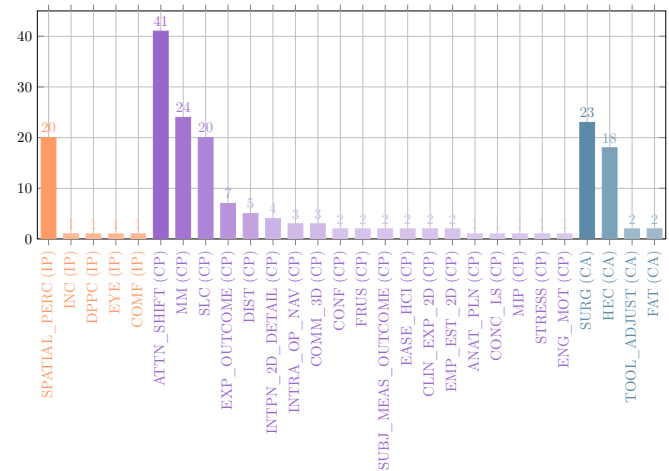

\subsection{Human factors of conventional surgery that are addressed by OST-HMD AR}
\label{section:human_factors_conventional_surgery}
Fig. \ref{fig:distribution_addressed_human_factors} shows the distribution of all human factors (out of the identified 34 ones) described as being a limitation of conventional non-AR surgical methods that the authors aimed to address with their proposed OST-HMD solution. We group these into the categories IP, CP and CA and describe the most popular in more detail.

\subsubsection{Information Perception related Human Factors of conventional surgery}

\textcolor{atomictangerine}{\textit{Spatial perception/awareness (SPATIAL\_PERC):}} A fundamental limitation of conventional image guidance methods appears to be the fact that crucial patient anatomy can only be perceived in 2D and hence prevents the surgeon from developing a personal sense of spatial perception and spatial awareness, which is a human factor we identified in (n = 20) articles and hence the dominating human factor of conventional surgery in the IP category. Impaired spatial awareness has the unwanted side effect of an increased likelihood of surgical errors due to misinterpretation of anatomical spatial relationships. \cite{qian2018arssist} aim to increase a first assistant's spatial awareness during robot-assisted laparoscopic surgery by providing a HoloLens solution in which a holographic endoscopy visualisation is registered within the personal viewing frustrum.
\cite{fotouhi2019interactive} addressed the problem of missing spatial context when looking at C-arm X-ray anatomy images on an external 2D monitor and presented a spatially aware HoloLens visualisation in which X-ray images are displayed in the correct spatial position of the patient's anatomy with a surgeon's view frustrum.

\subsubsection{Cognitive Processing related Human Factors of conventional surgery}
\label{section:cognitive_processing_human_factors_conventional_surgery}
\textcolor{amethyst}{\textit{Attention switch between surgical site and separate computer monitor (ATTN\_SHIFT):}} The human factor with the highest number of articles (n = 41) in the CP category (and the dominating factor accross all three categories IP, CP and CA) is the \emph{attention switch between the surgical site and a separate computer monitor}. In computer-assisted surgery a surgeon has to look away from the surgical site in order to see patient anatomy or surgical navigation information, and even switch between the surgical site and the screen multiple times during an operation. This inability to see both the surgical site and important patient anatomy or guidance information at the same time causes unwanted human behavior such as inconvenience and may also impact the continuity of the surgery \cite{chen2015development}. Especially during image-based surgical navigation, the surgeon has to constantly switch his attention while manipulating surgical navigation tools which comes with unwanted side effect such as unfamiliar hand-eye coordination, distraction and loss of concentration \cite{wang2016precision}. 

\textcolor{amethyst}{\textit{Mental Mapping of 2D image data to the 3D World (MM):}} An inherent problem of conventional computer-assisted surgery is that patient imaging data and surgical navigation information is displayed in 2D. This leads to the fact that the surgeon has to mentally map (or project) 2D image data onto the 3D world in order to translate the information seen on the 2D screen to the patient or surgical navigation tool. We identified \textit{mental mapping of 2D image data to the 3D world} in (n = 24) articles. \cite{Andres2018}, for example, addressed the problem of mental mapping in the context of intraoperative guidance in percutaneous orthopaedic surgical procedures in which a surgeon has to  place tools or implants precisely under C-arm based fluoroscopic imaging. The mental projection is counterintuitive and error-prone as a result of high mental workload and mental projective simplification. 

\textcolor{amethyst}{\textit{Steep learning curve (SLC):}} Some conventional surgical procedures, especially those related to image guidance, require surgeons to overcome a \textit{Steep Learning Curve} (n = 20) due to the method's inherent complexity. \cite{lin2018holoneedle} address this problem in needle guidance procedures that require considerable learning efforts due to the fact that physicians have to recover 3D information from 2D images, 
that the needle may cause artifacts in the images which hinder correct identification of needle tip and target and that complex hand-eye coordination
is required to register the 2D images seen on a separate monitor to the patient anatomy \cite{lin2018holoneedle}. Their OST-HMD AR system aims to reduce this learning curve.

\subsubsection{Control Action related Human Factors of conventional surgery}

\textcolor{airforceblue}{\textit{Increased risk of surgical error (SURG):}} Several researchers addressed the risk of surgical error (n = 23) which appears to be a common problem in some conventional image-guided procedures. 
\cite{el-hariri2018augmented} highlights the fact that conventional surgical navigation systems cannot observe the surgical scene and the external navigation computer monitor at the same time as being a potential problem that OST-HMD based solutions aim to solve. In another example, \cite{song2018endodontic} aim to prevent or reduce errors in root canal treatments such as accidental perforation during access cavity creation.

\textcolor{airforceblue}{\textit{Unfamiliar hand-eye coordination (HEC):}} The fact that a surgeon has to look away from the surgical site to a separate screen (see section \ref{section:cognitive_processing_human_factors_conventional_surgery}) while simultaneously manoeuvring surgical tools in image guided navigation causes unfamiliar hand-eye coordination because the surgeon cannot see his hands while looking on the separate screen \cite{wang2016precision}. \textit{Unfamiliar hand-eye coordination} is tackled in (n = 18) articles. \cite{qian2018arssist}, for example, addressed a first assistant's impaired hand-eye coordination during blind placement of robotic and hand-held instruments in conventional robot-assisted laparoscopic surgery by registering the holographic endoscopy visualisation with the visualised endoscope view frustrum  (see Fig. \ref{fig:qian2018arssist_fig4f_and_fotouhi2020reflective_fig4} (a) of section \ref{section:surgical_context}). Another example is given by \cite{deib2018image} who mention that a fundamental problem of conventional image guided percutaneous spine lies in an indirect guidance visualisation because radiography monitors showing fluoroscopic images are not aligned with the surgical site, which in turn hinders hand-eye coordination.

\subsection{Persistent human factors of proposed OST-HMD solutions}

\begin{figure}[ht!]
\resizebox{\columnwidth}{!}{
\begin{tikzpicture}
\begin{axis}[
    width=16.5cm,
    height=7cm,
    nodes near coords,
    xmin=-0.999 ,xmax=47,
    ymin=0, ymax=23,
    xtick=\empty,
    ybar=-\pgfplotbarwidth,
    extra x ticks={0, 2, 4, 6, 8, 10, 12, 14, 16, 18, 20, 22, 24, 26, 28, 30, 32, 34, 36, 38, 40, 42, 44, 46 },
    extra x tick labels={\textcolor{atomictangerine}{COMF (IP)}, \textcolor{atomictangerine}{SPATIAL\_PERC (IP)}, \textcolor{atomictangerine}{DPPC (IP)}, \textcolor{atomictangerine}{PER\_REAL\_AUG (IP)}, \textcolor{atomictangerine}{INC (IP)}, \textcolor{atomictangerine}{IMMR (IP)}, \textcolor{atomictangerine}{EYE (IP)}, \textcolor{atomictangerine}{PREF\_HOL (CP)}, \textcolor{amethyst}{EASE\_HCI (CP)}, \textcolor{amethyst}{USEF (CP)},  \textcolor{amethyst}{CONF (CP)}, \textcolor{amethyst}{FRUS (CP)}, \textcolor{amethyst}{SLC (CP)}, \textcolor{amethyst}{DIST (CP)}, \textcolor{amethyst}{SUBJ\_MEAS\_OUTCOME (CP)}, \textcolor{amethyst}{ANX (CP)}, \textcolor{amethyst}{CONC\_LS (CP)}, \textcolor{amethyst}{EXP\_OUTCOM (CP)}, \textcolor{amethyst}{COMM\_3D (CP)}, \textcolor{amethyst}{STRESS (CP)}, \textcolor{amethyst}{ENG\_MOT (CP)}, \textcolor{amethyst}{VIS\_OPT (CP)}, \textcolor{airforceblue}{FAT (CA)}, \textcolor{airforceblue}{SURG (CA)}},
    extra x tick style={
           tick label style={rotate=90} 
           },
    ymajorgrids,
    xmajorgrids,
    ]
\addplot[atomictangerine, fill] coordinates{(0, 15)}; 
\addplot[atomictangerine!90, fill] coordinates{(2, 14)}; 
\addplot[atomictangerine!70, fill] coordinates{(4, 7)}; 
\addplot[atomictangerine!60, fill] coordinates{(6, 4)}; 
\addplot[atomictangerine!40, fill] coordinates{(8, 1)}; 
\addplot[atomictangerine!40, fill] coordinates{(10, 1)}; 
\addplot[atomictangerine!40,fill] coordinates{(12, 1)}; 
\addplot[atomictangerine!40,fill] coordinates{(14, 1)}; 

\addplot[amethyst!100,fill] coordinates{(16, 20)}; 
\addplot[amethyst!80,fill] coordinates{(18, 7)}; 
\addplot[amethyst!60,fill] coordinates{(20, 4)}; 
\addplot[amethyst!60,fill] coordinates{(22, 3)}; 
\addplot[amethyst!60,fill] coordinates{(24, 2)}; 
\addplot[amethyst!60,fill] coordinates{(26, 2)}; 
\addplot[amethyst!60,fill] coordinates{(28, 2)}; 
\addplot[amethyst!60,fill] coordinates{(30, 2)}; 
\addplot[amethyst!40,fill] coordinates{(32, 1)}; 
\addplot[amethyst!40,fill] coordinates{(34, 1)}; 
\addplot[amethyst!40,fill] coordinates{(36, 1)}; 
\addplot[amethyst!40,fill] coordinates{(38, 1)}; 
\addplot[amethyst!40,fill] coordinates{(40, 1)}; 

\addplot[airforceblue,fill] coordinates{(42, 6)}; 
\addplot[airforceblue!80,fill] coordinates{(44, 3)}; 
\addplot[airforceblue!60,fill] coordinates{(46, 1)}; 
\end{axis}
\end{tikzpicture}
}
\caption{Distribution of persistent human factors of the proposed AR surgical approaches, grouped into the three categories \textcolor{atomictangerine}{1. Information Perception (IP)}, \textcolor{amethyst}{2. Cognitive Processing (CP)}, \textcolor{airforceblue}{3. Control Actions (CA)}}
\label{fig:distribution_persistent_human_factors}
\end{figure}
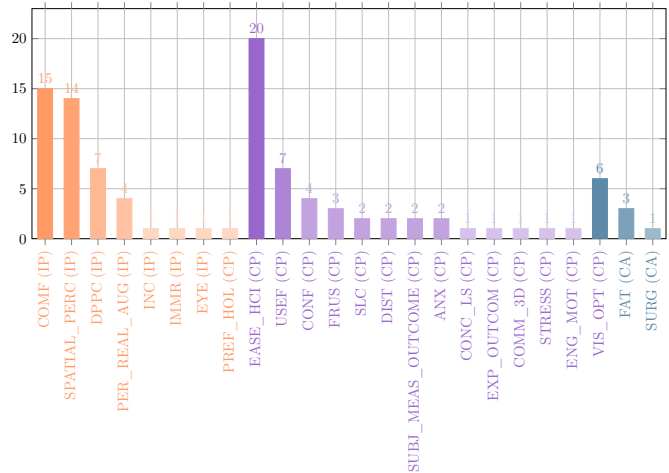

Since OST-HMDs expose the user to new and possibly unfamiliar visual perception and interpretation as well as interaction options, the proposed OST-HMD solutions also introduce new human factors that should be taken into account when designing effective HCI. We refer to these as \textit{persistent} human factors because they remain as issues of the proposed OST-HMD based solution. Table \ref{fig:distribution_persistent_human_factors} shows the distribution of persistent human factors, some of which we discuss in the following sections. Analogous to section \ref{section:human_factors_conventional_surgery}, the human factors are grouped into the categories IP, CP and CA and the most popular are detailed.

\subsubsection{Information Perception related Human Factors of AR-assisted surgery}

\textcolor{atomictangerine}{\textit{Perceived comfort level when wearing OST-HMD (COMF):}} As is the case for all HCI devices including computers or laptops, one of the most important factors that influence user acceptance is the comfort level. Discomfort will inevitably prevent a device from becoming a routine instrument that users enjoy working with. \textit{Perceived comfort level when wearing OST-HMD} was mentioned in (n = 15) articles, and is therefore one of the human factors that dominate the IP category. \cite{pietruski2019supporting} presented a Movierio BT-200 Smart Glasses based intraoperative navigation system that supports mandibular resection and conducted a phantom experiment in which osteotomies were performed. Surgeons reported good long term wear work ergonomics. \cite{rojas2020system} created a HoloLens telementoring system that allows surgeons to perform mentored leg fasciotomies. Participants reported that the weight of the HoloLens has a negative impact on their posture and comfort.

\textcolor{atomictangerine}{\textit{Spatial perception/awareness (SPATIAL\_PERC): }} Spatial perception and spatial awareness was already described in section \ref{section:human_factors_conventional_surgery} as a human factor of conventional non-AR methods, where 3D patient anatomy had to be inferred from 2D data. However, these spatial processing capabilities area also factors of 3D holographic visualisations and should also be taken into account for OST-HMD solutions, as reported in  (n = 14) articles.
Given that AR exposes users to new perceptual stimuli that are usually not part of their normal experience, it is likely that users processes this new visual information differently, which in turn impacts the quality of the HCI during OST-HMD assisted surgical procedures.
\cite{condino2018build} presented a HoloLens based hybrid simulator for orthopaedic open surgery that allows users to visualise 3D anatomy prior to performing a virtual viewfinder assisted surgical incision. Study participants who conducted a simulator experiment were engineers and clinicians. Results from a 5-point Likert questionnaire indicate that both user groups found it rather easy to perceive spatial relationships between real and virtual content; however, engineers tend to rate the ease of spatial relationship perception slightly higher than clinicians. 
\cite{gnanasegaram2020evaluating} investigated in how far ear anatomy learning can be improved compared to conventional didactic lectures and computer modules. Study participants performed a spatial exploration of holographic ear models displayed on a HoloLens and rated the OST-HMD higher than didactic lectures and computer modules in terms of 1.) overall learning effectiveness, 2.) the learning platform's ability to convey anatomic spatial relationships and 3.) learner engagement and motivation.

\textcolor{atomictangerine}{\textit{Missing/impaired depth perception (DPPC):}} Individual depth perception capabilities influence the ability to understand three dimensional holographic relationships as well as relationships between real and virtual objects. This may decrease the utility of systems that require perceptual precision. Several articles indicate missing or impaired depth perception as one of the limitations of the proposed OST-HMD approach (n = 7). 
\cite{Andres2018} developed a fluoroscopic X-ray guidance system for percutaneous orthopaedic surgery that is based on a cocalibration of a C-arm to a HoloLens and aims to facilitate the perception of spatial relationships between patient anatomy and surgical tools. A phantom based K-wire insertion experiment revealed that the HoloLen's build-in characteristic of rendering all holographic content at a focal distance of around 2m impacts the the user's depth perception, and hence leads to an impaired interaction between real and virtual objects.

\subsubsection{Cognitive Processing related Human Factors of AR-assisted surgery}

\textcolor{amethyst}{\textit{Perceived degree of ease and intuitiveness of HCI (EASE\_HCI):}}
A fundamental aspect that plays a pivotal role when it comes to user acceptance of a proposed OST-HMD solution is the \textit{perceived degree of ease and intuitiveness of HCI} which was the human factor with the most associated articles (n = 20) in the CP category. 
\cite{deib2018image} presented a HoloLens based application for image guided percutaneous spine procedures that was tested in a phantom experiment in which percutaneous vertebroplasty, kyphoplasty and discectomy interventions were performed. Participants could select their preferred holographic visualisation mode and questionnaire results revealed that 
initially the most popular mode was the option that was closest to a conventional 2D monitor and hence the most intuitive one. However, after the user became familiar with the OST-HMD environment, the preferred mode of visualisation changed to one that offers more benefits of the new mixed reality environment. \cite{jalaliniya2017wearable} designed a Google Glass based personal assistant for surgeons for which users who conducted experiments had to complete a questionnaire which revealed that some users prefer hand gesture interaction over voice interaction because voice interferes with their patient communication.

\textcolor{amethyst}{\textit{Perceived usefulness of OST-HMD (USEF):}} Since OST-HMDs are not well established in operating theatres and part of routine surgical procedures clinicians can always compare OST-HMD solutions with conventional methods and thus decide for themselves whether the new AR approach is useful or not. It is therefore not surprising that \textit{perceived usefulness} is a human factor mentioned in several articles (n = 7).
\cite{borgmann2016feasibility} conducted a feasibility of Google Glass assisted urological procedures in which surgeons could access holographic preoperative CT scans. A patient case study with a five-point Likert scale evaluation involving 7 surgeons over 10 procedures totalling 31 procedures revealed that the system's overall usefulness was rated as very high or high by 74\% of the surgeons.

\subsubsection{Control Action related Human Factors of AR-assisted surgery}

\textcolor{airforceblue}{\textit{Selection of preferred mode of visualisation (VIS\_OPT):}} Sometimes users are given the possibility to optimize their HCI experience by selecting one out of several different modes of visualization. 
The \textit{Selection of preferred mode of visualisation (VIS\_OPT)} considers personal optimisation of the HCI (n = 6) is the human factor with the most associated number of articles (n = 6) in the CA category. An example is described in \cite{qian2018arssist}: a first assistant has the option between two  modes of endoscopy visualization- during robotic surgery: 1.) A holographic monitor capturing the endoscopy camera stream or 2.) an endoscopy visualization that is registered with the viewing frustrum (Fig. \ref{fig:qian2018arssist_fig4f_and_fotouhi2020reflective_fig4} (a)).

\begin{table*}
\centering
\begin{threeparttable}
\caption{Addressed and persistent human factors (notation: human factor(s) on the left side of the arrow cause other human factor(s) on the right side of the arrow).}
\begin{scriptsize}
\rowcolors{2}{white}{lightgray!20}
\begin{tabular}{p{0.17\linewidth}p{0.4\linewidth}p{0.33\linewidth}}
\toprule
{\textbf{Study}} & {\textbf{Addressed Human Factors}} & {\textbf{Reported Persistent Human Factors}}\\
\midrule 
\cite{lin2018holoneedle}
& 
MM, INTPN\_2D\_DETAIL, HEC, SLC
&
N/A \\

\cite{qian2018arssist} 
&
ATTN\_SHIFT $\rightarrow$ HEC, TOOL\_ADJUST, DPPC, EXP\_OUTCOME, SPATIAL\_PERC
&
VIS\_OPT \\
\cite{chen2015development}
&
ATTN\_SHIFT $\rightarrow$ INC
&
N/A\\
\cite{wang2016precision}
&
[ATTN\_SHIFT $\rightarrow$ HEC, DIST, CONC\_LS], SLC
& 
N/A \\

\cite{deib2018image}
&
ATTN\_SHIFT $\rightarrow$ HEC
& 
EASE\_HCI, SLC, VIS\_OPT, COMF \\

\cite{Andres2018}
&
ATTN\_SHIFT $\rightarrow$ DIST, MM, SLC 
& 
DPPC \\

\cite{condino2018build}
&
SUBJ\_MEAS\_OUTCOME, SLC
& 
PER\_REAL\_AUG, FAT, IMMR, EASE\_HCI, SPATIAL\_PERC \\

\cite{stewart2016wearable}
&
ATTN\_SHIFT $\rightarrow$ DIST
& 
EYE \\

\cite{gibby2019head}
&
ATTN\_SHIFT $\rightarrow$ HEC
& 
VIS\_OPT \\

\cite{yoon2017technical}
&
ATTN\_SHIFT, HEC
& 
CONC\_LS, ANX \\

\cite{deOliveira2019}
&
ATTN\_SHIFT, SLC
&
N/A \\

\cite{aaskov2019x}
&
[INTPN\_2D\_DETAIL $\rightarrow$ SURG]
& 
PER\_REAL\_AUG \\

\cite{liebmann2019pedicle} 
&
INTRA\_OP\_NAV
& 
N/A \\

\cite{el-hariri2018augmented} 
&
[ATTN\_SHIFT $\rightarrow$ SURG\_ERR], [MM $\rightarrow$ SLC]
& 
N/A, DPPC \\

\cite{fotouhi2019interactive} 
&
[ATTN\_SHIFT $\rightarrow$ HEC], [MM $\rightarrow$ FRUS], SPATIAL\_PERC 
& 
N/A \\

\cite{meulstee2019toward}
&
ATTN\_SHIFT, MM
& 
N/A \\

\cite{song2018endodontic} 
&
[ATTN\_SHIFT $\rightarrow$ SLC \& SURG\_ERR], DPPC
& 
N/A \\

\cite{mitsuno2017intraoperative} 
&
ATTN\_SHIFT, SUBJ\_MEAS\_OUTCOME
& 
PREF\_HOL, PER\_REAL\_AUG \\

\cite{pratt2018through} 
&
[INTPN\_2D\_DETAIL $\rightarrow$ SURG\_ERR], DPPC
& 
N/A \\


\cite{brun2019mixed} 
&
DPPC, COMM\_3D, MM
& 
EASE\_HCI \\

\cite{li2017human} 
&
EMP\_EST\_2D, EASE\_HCI, CLIN\_EXP\_2D
& 
EASE\_HCI \\

\cite{zou2017coronary} 
&
EMP\_EST\_2D, EASE\_HCI
& 
EASE\_HCI \\

\cite{liu2019augmented} 
&
[DPPC, INTPN\_2D\_DETAIL $\rightarrow$ INTRA\_OP\_NAV]
& 
EASE\_HCI, VIS\_OPT \\

\cite{pietruski2019supporting} 
&
[ATTN\_SHIFT $\rightarrow$ HEC], DPPC, SPATIAL\_PERC
& 
COMF \\

\cite{kaneko2016ultrasound} 
&
[ATTN\_SHIFT $\rightarrow$ HEC]
& 
N/A \\

\cite{kuhlemann2017towards} 
&
MM 
& 
EASE\_HCI \\

\cite{karmonik2018augmented} 
&
COMM\_3D
& 
EASE\_HCI \\

\cite{frantz2018augmenting} 
&
MM, ATTN\_SHIFT
& 
SPATIAL\_PERC \\

\cite{fotouhi2020reflective} 
&
SPATIAL\_PERC, TOOL\_ADJUST, SLC 
& 
PER\_REAL\_AUG \\

\cite{hiranaka2017augmented}
&
[ATTN\_SHIFT $\rightarrow$ SURG\_ERR] 
&
N/A\\

\cite{katic2015system}
&
STRESS, MIP, [ATTN\_SHIFT $\rightarrow$ ERG, SURG\_ERR]
&
 VIS\_OPT \\

\cite{borgmann2016feasibility}
&
N/A
&
USEF \\

\cite{unberath2018augmented}
&
MM, INTRA\_OP\_NAV 
&
SPATIAL\_PERC, SUBJ\_MEAS\_OUTCOME  \\

\cite{sauer2017mixed}
&
MM, [ATTN\_SHIFT $\rightarrow$ HEC], SPATIAL\_PERC
&
COMM\_3D  \\

\cite{armstrong2014heads} 
&
SLC
&
N/A  \\

\cite{mahmood2018augmented}
&
SLC, MM, SPATIAL\_PERC
&
SPATIAL\_PERC, FAT  \\

\cite{rojas2019surgical} 
&
ATTN\_SHIFT, MM, FRUS, DPPC
&
FRUS \\

\cite{rojas2020system}
&
CLIN\_EXP\_2D, SLC, HEC, EXP\_OUTCOME
&
ANX, COMF, CONF \\

\cite{pelanis2020use}
&
MM, SPATIAL\_PERC 
&
SPATIAL\_PERC, COMF \\

\cite{nguyen2020augmented}
&
[ATTN\_SHIFT $\rightarrow$ SURG]
&
N/A \\

\cite{zhou2019design}
&
MM, SLC 
&
N/A \\

\cite{pietruski2020supporting}
&
MM, ATTN\_SHIFT
&
N/A \\

\cite{chien2019hololens}
&
ATTN\_SHIFT
&
N/A \\

\cite{zhang2019preliminary}
&
[ATTN\_SHIFT $\rightarrow$ MM, HEC], SPATIAL\_PERC 
&
COMF \\

\cite{heinrich2019holoinjection}
&
ATTN\_SHIFT, MM
&
DPPC \\

\cite{zhou2020surgical}
&
ATTN\_SHIFT, SURG 
&
SLC \\

\cite{wellens2019comparison} 
&
[ANAT\_PLN $\rightarrow$ SURG] 
&
SPATIAL\_PERC \\

\cite{fotouhi2019co}
&
MM 
&
SPATIAL\_PERC \\

\cite{baum2020augmented}
&
MM, EXP\_OUTCOME, SLC, SPATIAL\_PERC
&
SPATIAL\_PERC\\

\cite{liounakos2020head} 
&
[ATTN\_SHIFT $\rightarrow$ HEC]
&
COMF  \\

\cite{jalaliniya2017wearable}
&
ATTN\_SHIFT
&
EASE\_HCI  \\

\cite{rynio2019holographically}
&
SPATIAL\_PERC 
&
N/A  \\

\cite{boillat2019increasing} 
&
SURG
&
N/A \\

\cite{zhou2019towards}
&
SURG, ATTN\_SHIFT, HEC
&
DPPC  \\

\cite{schlosser2019exploratory}
&
DIST, FAT
&
DIST  \\

\cite{ponce2014emerging}
&
SLC
&
COMF  \\

\cite{guo2019online}
&
EYE, SUBJ\_MEAS\_OUTCOME
&
SUBJ\_MEAS\_OUTCOME \\

\cite{dickey2016augmented}
&
SLC, DIST
&
DIST, USEF, EASE\_HCI \\

\cite{al2020effectiveness}
&
[ATTN\_SHIFT, HEC $\rightarrow$ SURG, SPATIAL\_PERC]
&
COMF, SPATIAL\_PERC \\

\cite{li2019mixed}
&
SLC, ATTN\_SHIFT, HEC, SPATIAL\_PERC
&
N/A  \\

\cite{pepe2019marker}
&
SURG
&
N/A \\

\cite{wu2018augmented} 
&
ATTN\_SHIFT 
&
N/A \\

\cite{liebert2016novel}
&
ATTN\_SHIFT
&
N/A  \\

\cite{gnanasegaram2020evaluating}
&
ENG\_MOT
&
ENG\_MOT, SPATIAL\_PERC \\

\cite{sun2020high}
&
EXP\_OUTCOME
&
EASE\_HCI \\

\cite{park2020three}
&
ATTN\_SHIFT, SPATIAL\_PERC 
&
SPATIAL\_PERC \\

\cite{mendes2020pinata}
&
SLC
&
USEF, EASE\_HCI, FRUS \\

\cite{laguna2020assessing}
&
SPATIAL\_PERC, CONF
&
SPATIAL\_PERC \\

\cite{dallas2020comparing}
&
N/A
&
SPATIAL\_PERC \\

\cite{zafar2020evaluation}
&
SPATIAL\_PERC, CONF
&
EASE\_HCI, COMF, USEF \\

\cite{fitski2020mri}
&
DPPC, SPATIAL\_PERC
&
USEF, CONF \\

\cite{schoeb2020mixed}
&
N/A
&
EASE\_HCI, CONF, SLC \\

\cite{luzon2020value}
&
N/A
&
CONF, SPATIAL\_PERC \\

\cite{matsukawa2020smart}
&
[ATTN\_SHIFT $\rightarrow$ SURG, INC] 
&
N/A \\

\cite{yang2020feasibility}
&
SURG, SPATIAL\_PERC 
&
SURG \\

\multicolumn{3}{r}{(continued on next page)}
\end{tabular}
\end{scriptsize} 
\end{threeparttable}
\end{table*}
\begin{table*}
\centering
\begin{threeparttable}
\caption*{Table \thetable \hspace{0.2cm} (continued)}
\begin{scriptsize}
\rowcolors{2}{white}{lightgray!20}
\begin{tabular}{p{0.17\linewidth}p{0.4\linewidth}p{0.33\linewidth}}

\toprule
{\textbf{Study}} & {\textbf{Addressed Human Factors}} & {\textbf{Reported Persistent Human Factors}}\\
\midrule 

\cite{li2020smartphone}
&
[HEC $\rightarrow$ SURG]
&
FAT, INC, COMF \\

\cite{kumar2020use}
&
SPATIAL\_PERC, MM
&
DPPC, COMF \\

\cite{li2020clinical}
&
SPATIAL\_PERC, SURG
&
N/A \\

\cite{gibby2020use}
&
MM
&
EASE\_HCI \\

\cite{gu2020feasibility}
&
ATTN\_SHIFT, SURG
&
N/A \\

\cite{galati2020experimental}
&
ATTN\_SHIFT, SURG, MM, SLC
&
COMF, STRESS, EASE\_HCI, DPPC \\

\cite{viehofer2020augmented}
&
SLC, EXP\_OUTCOME, SURG
&
EXP\_OUTCOME \\

\cite{dennler2020augmented}
&
SLC, EXP\_OUTCOME, SURG, ATTN\_SHIFT 
&
N/A \\

\cite{kriechling2020augmented}
&
SURG, EXP\_OUTCOME
&
N/A \\

\cite{zorzal2020laparoscopy}
&
[ATTN\_SHIFT $\rightarrow$ HEC], SLC, COMF, FAT
&
DPPC, COMF, EASE\_HCI, USEF \\

\cite{cartucho2020multimodal}
&
N/A
&
EASE\_HCI, USEF, VIS\_OPT, COMF  \\

\cite{rojas2020evaluation}
&
[ATTN\_SHIFT $\rightarrow$ MM, SURG]
&
EASE\_HCI, FRUS  \\

\cite{scherl2020augmented}
&
N/A
&
EASE\_HCI, COMF \\

\cite{creighton2020early}
&
SPATIAL\_PERC
&
N/A \\

\cite{jiang2020hololens}
&
DPPC, MM
&
N/A \\

\cite{sun2020fast}
&
N/A
&
N/A \\

\bottomrule
\end{tabular}

\end{scriptsize}
\end{threeparttable}\end{table*}

\section{Discussion}
In this review we summarise the current proposed applications of OST-HMDs in surgery. Orthopaedic surgery applications are the most popular (30.16\%) and are mainly assisted intraoperative guidance applications, perhaps because it involves rigid bony structures and is a field where conventional guidance systems to achieve good implant alignment have become commonplace.

Image guidance, where a preoperative segmented imaging model is aligned to the operative view, is the dominating application across several surgical specialities. When providing such guidance and navigation, safety and accuracy becomes crucial. We summarised the achieved accuracy results in section \ref{section:accuracy} and noted that there is considerable variation in the reported results, which is related to large variation in terms of conducted experiments and different accuracy measures. Registration can be achieved manually or by identification of point or surface features using an external tracking device. There is no general solution to the problem of registration as yet.

The most common visualisation is of preoperative models. When these are generated from a preoperative scan this requires a segmentation process that must be incorporated into the surgical planning workflow. Medical image segmentation is a huge research area in its own right, with great progress being made. Though this is a vital component of image guidance, we have chosen not include it in this review of OST-HMD AR.

Beyond surgical guidance, other surgical application contexts where accuracy of superimposed holographic content may be less important have been analyzed in this review, such as preoperative planning or surgical training. Due to the variety of surgical contexts, different AR visualisations have been used, such as preoperative models, intraoperative images and intraoperative streaming of video, which all serve different purposes and are rated differently by users in terms of their usefulness.

Phantom experiments dominate, underlining the fact that many such systems are some way from clinical use. Aside from technological limitations, human factors have a major influence on the establishment of OST-HMD assisted applications in the operating room. Attention shift between the surgical site and an external computer monitor is the dominating human factor researchers aim to solve with OST-HMD solutions. These devices lead to other human factor issues, however, such as impaired hand-eye coordination and increased cognitive load that may increase rather than decrease the risk of surgical errors. 

\subsection{Human factor classification}
The presented human factors of this review reflect our attempt to identify human user's individual HCI characteristics in context of OST-HMD assisted surgery, providing an overview of perceptual and HCI related human characteristics that may impact the utility of a proposed novel AR-assisted system. Given that OST-HMD based surgical applications have not replaced respective conventional state of the art methods yet, we feel there is a need to increase awareness of all aspects that may influence the end user's acceptance of a novel technology being introduced in the operating room. Despite addressing several human factors, OST-HMD based solutions also expose the user to new human factors that may hinder an acceptance of this novel technology in the operating room.

The dominating persistent human factor is the perceived degree of ease and intuitiveness of HCI.
These new HCI possibilities may reveal individual performance differences and user preferences even more than conventional computer assisted surgical methods. Overall, it appears that the combination of OST-HMD device, surgical speciality, surgical application context, surgical procedure, proposed AR visualisation and conducted experiments triggers different individual human HCI responses that lead to variation in individual perceived utility. 
Some attempts have been made to provide standardised analysis in image guidance applications. \cite{zuo2020novel} proposed a novel multi-indicator evaluation model for mixed reality surgical navigation systems that evaluate the user's perception in regards to safety, comfort and efficiency and combines subjective and objective evaluation criteria. \cite{doswell2014augmenting} identified the need for HMD-based, scientifically grounded methods that identify HCI related interaction modalities to be addressed to optimise user performance and cognitive load. These comprise information presentation, user input and system feedback. the suggest that an ideal HCI system should be able to adapt these interaction modalities in real-time and in response to the given task as well as environmental and user psychophysiological states.

A taxonomy for mixed reality visualisation in image guided surgery has been proposed by \cite{kersten2012dvv} aiming to introduce a new common framework that facilitates the establishment of validation criteria and should lead to more mixed reality systems being used in daily surgical routine. The paper is well cited and the comprehensive literature review of AR in laparoscopic surgery from \cite{bernhardt2017status} categorises articles according to their taxonomy. But the translation  into commercial applications that are used on a daily basis in operating rooms has not materialised as yet. A similar taxonomy tailored to OST-HMD AR would be desirable but is hard to achieve given the widely varying needs of the implementations presented in this review.

\subsection{Potential machine learning applications}
Given the increasing trend of machine learning (ML) applications for medical image processing, such methods are likely to be applied to OST-HMD solutions. However, none of the selected 91 articles contained such a ML application. OST-HMD systems provide a 3D world with a wealth of data, including video images, gesture-based interaction data, eye tracking and generated surface meshes. These could provide rich training data for ML algorithms. 

ML algorithms have been proposed for surgical mixed reality applications. \cite{azimi2018interactive} presented an interactive training and operation ecosystem for mixed reality related surgical tasks that includes data collection for potential ML algorithms. Their system records data from multiple users, such as gaze tracking to indicate which locations in 3D space a surgeon is paying attention to. ML algorithms could then use this data to identify novice surgeons and activate guidance support.

Another example aiming to expand the user's hands-free interaction possibilities when wearing a HMD was proposed by \cite{chen2019gaze}. Using a self-made HMD with eye-tracking cameras, the authors proposed a deep convolutional neural network to classify gaze trajectories and gaze trajectory gestures. The classified gestures in turn can then trigger different HCI operations.
The HoloLens 2 comes with built-in gaze tracking and offers new HCI possibilities that still need to be explored, especially in a surgical setup. A fundamental step in terms of accelerating ML research in the surgical field will be the creation of data bases with relevant user data, which can then serve as inputs for ML algorithms. 

\subsection{Conclusions}

The field of OST-HMD assisted surgery has shown a significant recent upward trend in the number of publications as well as the diversity of surgical applications that could benefit from this technology. The release of the Microsoft HoloLens has boosted research into mixed reality surgical applications from 2017 onwards (see table \ref{table:hmd_surgical_phase_surgical_applications}). However, comparatively few systems have been used clinically to date and demonstration of utility is rare. 

It is worth noting that \cite{dilley2019perfect}, in a screen-based simulation system, compared direct AR with nearby unregistered guidance. Overlaid AR was found to cause inattention blindness, where the augmented view distracts from important events in the real view. This problem arises even when registration is perfect. One option is that guidance information could be presented near to, but not overlaid directly on the surgical view. Such a side-by-side visualisation would allow correctly oriented, but not fully registered model data to be readily available without obscuring or confusing the real view.

The training aspect of OST-HMD visualisation should not be underestimated. The ability to view 3D anatomy and pathology in situ may improve spatial understanding in novice surgeons and reduce the learning curve. \cite{louis2020high} demonstrated improved learning with AR under high fidelity conditions. There is a case for similar experiments to be conducted using OST-HMD AR to provide evidence of proven benefit to learning with these devices.

One potential direction for research is a human factors approach that starts by identifying explicit points or moments in a procedure that may affect patient outcome and then tailors visualisations to improve performance and decision-making for these specific tasks. Demonstration of improved performance on similar specific tasks in the laboratory setting might also lead to a better understanding of the optimal role of AR.

Increasing exposure to AR devices may also improve acceptability of such technology. Taking both technological and human factors into consideration from the outset should lead research towards effective clinical implementations that finally realise the full potential of surgical AR.

\section{Acknowledgements}
The work was supported by the Wellcome/EPSRC Centre for Interventional and Surgical Sciences (WEISS) [203145Z/16/Z]; Engineering and Physical Sciences Research Council (EPSRC) [EP/P027938/1, EP/R004080/1, EP/P012841/1]; The Royal Academy of Engineering [CiET1819/2/36].

\begin{appendices}
\section{Papers summary table}
\setcounter{table}{0}
\renewcommand{\thetable}{\Alph{section}\arabic{table}}
\newcolumntype{C}[1]{>{\left\arraybackslash}p{#1}}
\onecolumn
\begin{landscape}
\pagestyle{empty}
\setlength{\hoffset}{-1.5cm}
\begin{table*}
\centering
\begin{threeparttable}
\caption{Description of AR visualization, Conducted experiments and accuracy of final 91 articles used for quantitative synthesis. Acronyms: PM: Preoperative model; II: Intraoperative image. PI: Preoperative image. IM: Intraoperative model. IV: Intraoperative live streaming video. PV: Preoperatively recorded video. DOC: Documents. COMM: 2D plane with video communication software application (google hangouts etc.). IND: Intraoperative numerical data. SSE: System setup experiment without phanton, cadaver or patient involvement (may contain additional hardware). PE: Phantom experiment. HCE: Human cadaver experiment. AE: Animal experiment. ACE: Animal cadaver experiment. SE: Simulator experiment. SCE: Simulated clinical environment experiment. PS: Patient case study.
Abbreviations:  Quan: Quantitative study. Qual: Qualitative study.} \label{table:appendix}
\begin{scriptsize}
\rowcolors{2}{white}{lightgray!20}
\begin{tabular}{p{0.12\linewidth}p{0.27\linewidth}p{0.27\linewidth}p{0.28\linewidth}}
 \toprule 
     \textbf{Study}  & \textbf{AR visualizations} & \textbf{Experiments} &  \textbf{Reported Accuracy}\\
  \midrule
 
  \cite{chen2015development} 
  & 
  PM: optimal bone drill trajectory, organs, bone structures
  & 
  Quan: PE: 1.) registration accuracy, 2.) surgical navigation. HCE: 3.) joint screw implantation
  
& 1.) $0.809 \pm 0.05$ $mm$, $1.038^{\circ}\pm0.05^{\circ}$
.  
\\ 
  
  \cite{wang2016precision}&
  PM: 3D pelvis model incl. vessels, optimal bone drill trajectory
  & 
  Quan: HCE: joint screw implantation
  &  $2.7\pm1.2$ $mm$, $3.7\pm1.1 mm$, $2.9^{\circ}\pm1.1^{\circ}$ 
  \\
  
  \cite{deib2018image} 
  & 
  II: radiographic images
  &
  Quan: \& Qual: PE: Percutaneous vertebroplasty, kyphoplasty and discectomy interventions
  & 
  N/A
   \\
  
  \cite{Andres2018} 
  & II: 2D X-ray images inkl. annotations, IM: guiding lines, planes \& spheres, C-arm source position (cylinder)
   & 
   Qual: SSE: 1.) Calibration, 2.) HMD tracking, 3.)Landmark identification, PE: 4.) K-wire guidance, 5.) Entry point localization (implantation of nails)
& 
1.) $21.4\pm11.4$ $mm$, 2.) $16.2\pm9.5$ $mm$, 3.) $8.76-11.7\pm(3.21-4.03)$ $mm$, 4.) $4.47\pm2.91$ \$ $9.84\pm3.97$, 5.) $5.2$ $mm$
\\
  
  \cite{condino2018build}
  & 
  PM: Anatomical 3D models (incl. bones \& muscles), virtual menu with toggle buttons, preoperative plan. IM: optimal tool trajectory
  & 
  Quan: 1.) System accuracy estimation (perceived AR target positions). 
  Qual: 2.) Subjective workload assessments (NASA Task Load Index)
  & 
1.) $0.6$ $mm$
  \\
  
  \cite{stewart2016wearable}
  & 
  IM: pose of surgical tool (stack of cyan
rings), navigation target (circle)
  & 
  Quan: PE: 1.) tracked tool positioning \& orienting, Qual: 2.) questionnaire
  & 
  1.) $ 0.40$ $\pm$ $0.78$ $mm$, $2.07$ $\pm$ $1.68^{\circ}$
 \\
  
  \cite{gibby2019head}
  & 
  PM: virtual trajectories (pedicle screw guidance), lumbar spine 2D \& 3D CT images
  & 
  Quan: PE: 1.) Registration accuracy verification, 
  2.) Percutaneous placement
  & 
  1.) $12.99$ mm ($12.31 – 13.61$ mm), 2.) $12.99$ mm ($12.31 – 13.61$ mm), $15.59$ mm ($12.02 – 18.69$ mm)
  \\
  \cite{deOliveira2019}
  & 
  PM: 3D organs incl. fiducial or anatomical markers
& 
Quan: 1.) reliability assessment of 
virtual-physical mappings, Quan: \& Qual: 2.) assessment of
superimposed holograms in physical space
&
1.) $19.74 \pm 2.38 mm$ (x), $76.82 \pm 3.83 mm$ (y), $−2.74 \pm 1.96 mm$ (z), 
$19.74 \pm 2.38^{\circ}$, $76.82 \pm 3.83$, $−2.74^{\circ} \pm 1.96^{\circ}$, 
2.) $3.2 \pm 1.6$ mm (RMSE)
\\
\cite{aaskov2019x}
&
PI: anteroposterior lumbar X-ray 2D images
& 
Qual: PS: 1.) Accuracy and 2.) repeatability validation
& 
1.) $8.77$ mm
\\
\cite{liebmann2019pedicle}
&
PM: targeted screw trajectory, drill entry points. IM: drill angle between current and targeted screw trajectory,  3D trajectory deviation triangle
& 
Quan: PE: guiding wire placement for pedicle screw
& 
$2.77 \pm 1.46$ mm, $3.38^{\circ} \pm 1.73^{\circ}$ 
\\
\cite{fotouhi2019interactive}
& 
II: interventional X-ray images, IM: view frustrum
& 
Quan: SSE: 1.) Hand-eye calibration experiment, PE: 2.) internal fixation of pelvic ring fractures \& percutaneous vertebroplasty 
& 
1.) $0.43$ mm $\pm$ $0.34$ mm, $0.43^\circ$ $\pm$ $0.34^\circ$
\\

\cite{lin2018holoneedle}
&
IM: needle visualizations (needle position, orientation \& shape, tangential ray)
& Quan \& Qual: PE: needle insertion task
& 
$8.15$ mm $\pm$ $0.4$ mm, $6.54$ mm $\pm$ $0.294$ mm, $6.03$ mm $\pm$ $0.291$ mm
\\

\cite{meulstee2019toward}
& 
PM: 3D objects (cube)
 & Quan: PE: 1.) Tight-Fit \& Loose-Fit Accuracy Evaluation
& 
1.) $0.7$ $\pm$ $0.2$ mm, 2.) $2.3$ $\pm$ $0.5$ mm
\\

\cite{guo2019online}
&
3D calibration cubes
& 
Quan: SSE: calibration accuracy
& 
below $6$ mm, up to $5^\circ$
\\

\cite{brun2019mixed} 
&
PM: 3D heart models
& 
Quan: SSE: patient based heart model analysis (anatomy identification \& diagnosis)
& 
N/A
\\

\cite{li2017human}
&
PM: 3D coronay arteries models
& 
Quan: SSE: Dynamic and static gesture recognition
&
N/A
\\ 

\cite{zou2017coronary} 
& 
PM: 3D cardio artery vascular models
& 
Quan: SSE: hand gesture recognition rate validation
& 
N/A
\\

\cite{liu2019augmented}
& 
PM: 3D heart, spine \& cathether models, 3D catheter path planning
& 
Quan: PE: catheter navigation under C-arm fluoroscopy guidance
&
$0.425\pm 0.021$ mm (registration), $ 0.29\pm 0.19$ mm (catheter position)
\\

\cite{kaneko2016ultrasound}
& 
II: ultrasound images
& 
Quan: SE: sonographic guided jugular vein catheterization
& 
N/A
\\

\cite{kuhlemann2017towards}
& 
PM: 3D patient surface mesh, vascular tree, catheter position, registration landmarks, canvasses (1.) 2D CT slide, 2.) catheter point of view perspective inside vascular tree)
& 
Quan: PE: 1.) calibration, 2.) catheter insertion \& navigation, 3.) Likert scale questionnaire evaluation
& 
1.) 1.) $4.34\pm 0.709$ mm (RMSE point-to-point correspondence)
\\

\cite{karmonik2018augmented}
&
PM: complex medical vascular \& blood flow 3D image data 
& 
Quan: SSE: Evaluation of vascular \& blood flow image data
& 
N/A
\\

\cite{frantz2018augmenting}
& 
PM: 3D skull visualizations, localization markers
& 
Quan: PE: 1.) Manual registration, 2.) Maintaining hologram registration via continuous camera tracking
& 
1.) $4.39$ $\pm$ $1.29$ mm, 2.) $1.4$ $\pm$ $0.67$ mm (mean perceived drift)
\\

\cite{yoon2017technical}
&
II: 2D neuronavigation images
& 
Quan: PS: pedicle screw placement
& 
N/A
\\

\multicolumn{4}{r}{(continued on next page)}
\end{tabular}
\end{scriptsize} 
\end{threeparttable}
\end{table*}
\begin{table*}
\centering
\begin{threeparttable}
\caption*{Table \thetable \hspace{0.2cm} (continued)}
\begin{scriptsize}
\rowcolors{2}{white}{lightgray!20}
\begin{tabular}{p{0.12\linewidth}p{0.27\linewidth}p{0.27\linewidth}p{0.28\linewidth}}
 \toprule 

     \textbf{Study}  & \textbf{AR visualizations} & \textbf{Experiments} &  \textbf{Reported Accuracy}\\
  \midrule

\cite{pietruski2019supporting}
& 
II: 2D navigation monitor. PM: 3D mandible model, 3D osteotomy cutting guides (planes) \& navigated surgical saw, IND: cutting guide deviation coordinate system
&
Quan: PE: osteotomies: 1.) augmented navigation system monitor \& 2.) superimposition of surgical plan
&  
1.) $1.79$ $\pm$ $0.94$ mm,  $3.67$ $\pm$ $3.67^{\circ}$, 2.) $2.41$ $\pm$ $1.34$ mm,  $7.14$ $\pm$ $5.19^{\circ}$
\\

\cite{mitsuno2017intraoperative}
& 
PM: Preoperative \& ideal postoperative 3D facial surface and facial bones
& 
Quan: PS: reconstructive surgeries (facial fractures or deformities)
&
$30 - 40$ mm (display error)
\\

\cite{pratt2018through}
& 
PM: 3D bony, vascular, skin \& soft tissue structures, vascular perforators, bounding box
& 
Quan: PS: flap surgery
& 
N/A
\\

\cite{fotouhi2020reflective}
& 
PM: 3D virtual robot arm, 2D reflective AR display 
& 
Quan: PE: Registration with 1.) and without 2.) reflective AR displays, 2.) Simulated robot-assisted trocar placement
& 
1.) $ 16.5  \pm  11.0 $ mm, 2.) $30.2 \pm 23.9$ mm (misalignment error)
\\

\cite{qian2018arssist}
& 
PM: \& II: 3D plane with endoscopy visualization, IM: viewing frustrum, PM: endoscope, robotic \& hand-held instruments
& 
Quan: SSE: 1.) Display calibration, 2.) Camera calibration. PE: Visualization performance evaluation
& 
1.) $4.27 \pm 3.09$ mm 
\\

\cite{song2018endodontic}
& 
PI: 2D radiographic images with guidance information, IM: 3D drill guidance information 
& 
Quan: PE: 1.) Accuracy evaluation, 2.) Tool navigation \& guidance 
& 
1.) Avg: $0.46$ mm, Max: $0.86$ mm, Avg: $1.17^{\circ}$ , Max: $2.10^{\circ}$ 
\\ 

\cite{hiranaka2017augmented} 
&
II: fluoroscopic video
&
Quan: PE: Guide wire insertion into femur
&
$2.6 \pm 0.02$ mm \\ 

\cite{katic2015system}
&
PM: \& IM: position, depth \& alignment of planed \& actual dental drill, injury avoidance warnings, drill heads
&
Quan: 1.) SSE: Calibration accuracy,  2.)  ACE:  Implant placement
& 
1.) $3.01 \pm 3.01$ mm, 2.) $< 2.5$ mm (implant deviation),  \\  

\cite{borgmann2016feasibility}
&
II: preoperative CT scan
&
Quan: PS: different urological procedures, Likert scale questionnaire
&
N/A \\

\cite{unberath2018augmented}
&
IM: Live 3D point cloud (C-arm pose)
&
Quan: PE: pelvic trauma surgery 
&
$51.6 \pm 19.2$ mm, $1.54 \pm 0.92^{\circ}$\\  

\cite{sauer2017mixed}
&
PM: 3D hepatic artery, portal vein, hepatic veins, liver tumor, liver capsule
&
Quan: PS: open hepatic surgery
&
N/A \\  

\cite{armstrong2014heads} 
&
COMM: google hangouts, DOC: articles from senior author
&
Quan: PS: reconstructive limb salvage procedure
&
N/A \\  
 
\cite{mahmood2018augmented}
&
PM: 3D anatomical models, 3D ultrasound streaming plane
&
Quan: SE: transesophageal echocardiography 
&
N/A \\  

\cite{rojas2019surgical}
& 
PM: 3D graphical annotations (incision lines, surgical instruments)
&
Quan: SE: 1.) anatomical marker placement, 2.) mock abdominal incision
&
1.) $ 11.37 \pm 0.72$ mm \\  

\cite{li2019mixed}
&
PM: 3D liver structure (intraoperatively updated), tumor, virtual needle, registration landmarks
&
Quan: 1.) PE: Registration accuracy validation, 2.) AE: needle insertion operation
&
1.) $2.24$ mm (avg. target registration error) \\  

\cite{rojas2020system}
&
PM: 3D graphical annotations (lines \& models)
&
Quan: HCE: leg fasciotomy
&
N/A. \\  

\cite{pelanis2020use}
&
PM: 3D liver incl. parenchyma, portal, hepatic veins \& lesion
&
Quan: SSE: Identification of liver segments
&
N/A \\  

\cite{pepe2019marker}
&
PM: 3D landmarks, 3D tumors, 3D axial facial CT slice
&
Quan: PE: Automatic registration after user calibration
&
 x: $ 3.3 \pm 2.3$ mm \,  y: -$4.5 \pm 2.9$ mm \,  z: -$9.3 \pm 6.1$ mm \\  

\cite{nguyen2020augmented}
&
PM: 3D patient head incl. skin, skull \& spine
&
Quan: PE: Registration accuracy. 3 registration methods: 1.) Keyboard, 2.) Tap to Place, 3.) 3-Point correspondence matching
&

1.) X Axis: $5 \pm 5^{\circ}$ \,  Y Axis: -$5.9 \pm 5.9^{\circ}$ \,  Z Axis: $6.8 \pm 5.9^{\circ}$; displacement: XY Plane: $2.9 \pm 1.8$ mm \,  ZY Plane: $1.8 \pm 1.2$ mm \,  XZ Plane: $1.6 \pm 0.9$ mm   \\  


\cite{zhou2019design}
&
PM: 3D organs, needle (actual \& preoperative plan)
&
Quan: 1.) PE: \& 2.) AE: needle insertion 
&
1.) $0.664$ mm, $4.74^{\circ}$, 2.) $1.617 $ mm, $5.574^{\circ}$ \\  

\cite{pietruski2020supporting}
&
PM: 3D bones, surgical plan: control points, osteotomy trajectories, navigated saw, 2D digital coordinate system
&
Quan: PE: osteotomy
&
$4.1 \pm 2.29$ mm, $5.08 \pm 3.64^{\circ}$, $4.97 \pm 2.91^{\circ}$ \\  

\cite{chien2019hololens}
&
PM: 3D patient skin surface
&
Quan: PE: Alignment (different data sparsity percentages are tested but we refer only to 100 \% of floating data being used)
&
5 reference points alignment error RMSE: Avg.: $0.932$ mm, Min: $0.37$ mm, Max: $1.49$ mm    \\

\cite{zhang2019preliminary}
&
PM: 3D intracranial structure, lesion
&
Quan: PS: Craniotomy
&
N/A. \\  

\cite{heinrich2019holoinjection}
&
PM: 3D Needle insertion guidance visualization options: 1.) planes,  2.) lines, 3.) cone rings
&
Quan: PE: 1.) Registration accuracy estimation: a) Angle measurement of displayed lines I, b) Angle measurement of displayed lines II, c) Tracked normal vector accuracy, d) Tracked normal vector accuracy, 2.)  Comparison study
&
1 a.) $0.76 \pm 0.11^{\circ}$, 1 b.) frontal viewing pos. $1.90 \pm 1.82^{\circ}$, $45^{\circ}$ viewing pos. $4.28 \pm 4.09^{\circ}$, lateral viewing pos. $7.94 \pm 7.75^{\circ}$, 1 c.) $0.72 \pm 0.41^{\circ}$, 1 d.) X \$ Y marker: $0.27 \pm 0.21^{\circ}$, X \$ Z marker: $0.31 \pm 0.22^{\circ}$, Y \$ Z marker: $0.38 \pm 0.36^{\circ}$ \\ 

\cite{zhou2020surgical}
&
3D anatomy (skin, bones, tumor tissue), virtual needles (planning \& detected), seeds, 2D control panel
&
Quan: 1.) PE: \& 2.) AE: brachytherapy of tumors 
&
Avg. needle location error: 1.) $0.957$ mm, 2.) $2.416 $ mm\\ 

\cite{wellens2019comparison} 
&
PM: 3D kidneys incl. tumor, arteries, veins, urinary collecting structures
&
Quan: SSE: Assessment of anatomical structures
&
N/A \\

\multicolumn{4}{r}{(continued on next page)}
\end{tabular}
\end{scriptsize} 
\end{threeparttable}
\end{table*}
\begin{table*}
\centering
\begin{threeparttable}
\caption*{Table \thetable \hspace{0.2cm} (continued)}
\begin{scriptsize}
\rowcolors{2}{white}{lightgray!20}
\begin{tabular}{p{0.12\linewidth}p{0.27\linewidth}p{0.27\linewidth}p{0.28\linewidth}}
 \toprule 
     \textbf{Study}  & \textbf{AR visualizations} & \textbf{Experiments} &  \textbf{Reported Accuracy}\\
  \midrule

\cite{fotouhi2019co}
&
IM: 3D anatomical structures, C-arm principle axis
&
Quan: SSE: 1.) calibration accuracy, PE: 2.) Target augmentation error, 3.) Augmented surgical visualization 
&
1.) $5.7 \pm 0.26$ mm, 2.) $10.8 \pm 3.45$ mm \\ 

\cite{baum2020augmented}
&
PM: 3D patient skin surface, brain, intra-cortical lesion
&
Quan: PE: Target Localization
&
N/A \\ 

\cite{liounakos2020head} 
&
II: live endoscopic camera image
&
Quan: PS: Lumbar discectomy
&
N/A \\ 

\cite{jalaliniya2017wearable}
&
COMM: videoconferencing application, PI: patient records
&
Quan: PE: SCE: Mobile access to patient records, telepresence
&
N/A \\ 

\cite{rynio2019holographically}
&
PM: 3D arterial system, aneurysm, bones, PI: 2D image with volume rendering, arterial diameters \& planning notes
&
Quan: PS: Abdominal aortic aneurysm repair
&
N/A \\ 

\cite{boillat2019increasing}
&
PM: 2D surgical safety checklist
&
Quan: SSE: time-out checklist execution
&
N/A \\ 

\cite{zhou2019towards}
&
PM: 3D tooth, cone (endoscope view frustrum), probe alignment cyclinder \& planes, II: 2D imaging
&
Quan: PE: 1.) Augmentation quality evaluation, 2.) Dental decay localization
&
$31 \pm 11$ px (keypoint displacement) \\ 

\cite{schlosser2019exploratory}
&
IND: 2D screen incl. patient heart rate, blood pressure, blood oxygen saturation, alarm notifications 
&
Quan: \& Qual: PS: vital sign monitoring, Quan: situation awareness measurement
&
N/A \\ 

\cite{ponce2014emerging}
&
IV: hybrid image (surgical field combined with hands of remote surgeon)
&
Quan: PS: shoulder replacement
&
N/A \\ 

\cite{dickey2016augmented}
&
IV: interactive video display incl. cursor moved by supervising physician, PV: training guide
&
Qual: \& Quan: SSE: user survey
&
N/A \\ 

\cite{al2020effectiveness}
&
PI: CT images, IV: live fluoroscopy, endoscopic view
&
Qual: \& Quan: SE: mid-ureteric stone removal
&
N/A \\

\cite{wu2018augmented}
&
PM: 3D patient anatomy (e.g. head, intracranial vascular tissue)
&
Quan: PE: dummy head alignment test
&
$< 3$ mm (Avg. Target Registration Error)\\

\cite{el-hariri2018augmented}
&
PM: 3D bone structures, fiducial markers
&
Quan: PE: accuracy assessment
&
Fiducual marker comparisons (RMSE): x: $3.22$ mm, y: $22.46$ mm, z: $28.30$ mm \\

\cite{liebert2016novel}
&
IND: 2D screen incl. patient arterial line blood pressure,
heart rate, heart rhythm, pulse oximetry, respiratory rate
&
Quan: SE: Vital signs monitoring during bronchoscopy
&
N/A \\ 

\cite{gnanasegaram2020evaluating}
&
PM: 3D ear anatomy
&
Quan: SSE: spatial exploration of holographic ear model
&
N/A \\

\cite{sun2020high}
&
PM: 3D catheter
&
Quan: SSE: 1.) Stability measurement of tracking algorithm, 2.) testing of tracking accuracy 
3.) latency test using third-party tracker, 4.) HCE: EDV performed on a cadaveric head
&
2.) avg. distance from catheter tip to corresponding grid intersections (2D plane): $0.58$ mm, overall avg. accuracy on all 3
grid faces: 0.85 mm (3D space) \\

\cite{park2020three}
&
PM: 3D volumes from MRI images
&
Quan: AE: transarterial embolization of hepatocellular carcinoma (HCC)
&
N/A \\

\cite{mendes2020pinata}
&
PM: 3D model of body simulator's external surface (upper torso) and 3D vascular structures
&
Quan: SE: tracked needle insertion
&
N/A \\

\cite{laguna2020assessing}
&
PM: 3D elbow fractures (bones)
&
Quan: SSE: Orthopedic surgeons' assessment of 3D AR models for presurgical planning in complex pediatric elbow fractures
&
N/A \\

\cite{dallas2020comparing}
&
PM: 3D spine model with a vascular model overlay
&
Quan: SSE: 3D model measurement using circumference
and angle tools of standard-of-care PACS software
&
N/A \\

\cite{zafar2020evaluation}
&
PM: 3D human skull
&
Quan: SSE: digital anatomy session with the HoloHuman virtual anatomy training software
&
N/A \\

\cite{fitski2020mri}
&
PM: 3D intraparenchymal arteries and veins, kidney, tumor
&
Quan: PS: Preoperative planning of patients eligible for nephron-sparing surgery (NSS)
&
N/A \\

\cite{schoeb2020mixed}
&
PV: 2D plane with catheter placement instruction guidance
&
Quan: SE: Bladder catheter placement using a male catheterization-training model
&
N/A \\

\cite{luzon2020value}
&
PM: 3D anatomy models (e.g. vascular model)
&
Quan: PE: registration and needle placement 
&
Target error distance: x-axis: $2.9757 \pm 1.33396$ mm, y-axis: $2.2790 \pm 1.44992$ mm, z-axis: $2.7844 \pm .91323$mm \\

\cite{matsukawa2020smart}
&
II: fluoroscopic 2D image
&
Quan: PS: single-segment posterior lumbar interbody fusion (PLIF) at L5–S1
&
N/A \\

\cite{yang2020feasibility}
&
PM: 3D portal vein and hepatic vein, liver
&
Quan: 1.) AE: dogs: simulated percutaneous puncture of the portal vein and
simulated TIPS, 2.) PE: liver phantom experiment
&
N/A \\

\cite{li2020smartphone}
&
PM: 3D internal organs, PI:  CT images,  IM: progress view of the virtual planned target,  needle path, skin entry point and needle end
&
Quan: PE: 1.) image overlay accuracy using 3D abdominal phantom, 2.) needle placement performance using tissue phantom
&
1.) Total target overlay error over 336 targets: $1.74 \pm 0.86$ mm. Needle overlay angle: $0.41 \pm 0.23^{\circ}$ \\

\multicolumn{4}{r}{(continued on next page)}
\end{tabular}
\end{scriptsize}
\end{threeparttable}
\end{table*}

\begin{table*}
\centering
\ContinuedFloat
\begin{threeparttable}
\caption*{Table \thetable \hspace{0.2cm} (continued)}
\begin{scriptsize}
\rowcolors{2}{white}{lightgray!20}
\begin{tabular}{p{0.12\linewidth}p{0.27\linewidth}p{0.27\linewidth}p{0.28\linewidth}}
 \toprule 
     \textbf{Study}  & \textbf{AR visualizations} & \textbf{Experiments} &  \textbf{Reported Accuracy}\\
  \midrule

\cite{kumar2020use}
&
PM: 3D liver and heart, slicing tool (plane)
&
SSE: visualisation of patient-specific models and hologram interaction (rotate, scale and move)
&
N/A \\

\cite{li2020clinical}
&
PM: 3D target organs and tumors (kidney, tumor, renal vessels, renal collection system, skin, skeleton, liver, spleen), PI: MR results, IV: laparoscopic video stream
&
Quan: PS: Prospective review of patients with stage T1N0M0 renal tumors who untervent laparoscopic partial ephrectomy 
&
N/A \\

\cite{gibby2020use}
&
PI: 3D plane with axial CT image, PM: needle trajectories in correct spatial orientation over patient 
&
Quan: 1.) PE: control data experiment (needle navigation) using skull with ballistic gelatin and radiopaque balls (targets), 2.) PS: interventional spine procedures 
&
1.) $0.998 \pm 1.66$ mm (mean error of needle tip to targeted ball). Mean distance from model surface to targeted ball: $79.42 \pm 15.33$ mm, 2.) Mean error of needle to target: $1.73 \pm 2.20$ mm \\

\cite{gu2020feasibility}
&
PM: 3D glenoid and planed drilling path
&
Quan: PE: 1.) inside-out registration (via HoloLens depth sensing camera), 2.) accuracy evaluation of inside-out registration using outside-in tracking with optical tracker,  3.) registration with surface digitisation
&
1.) Inside-out registration accuracy compared with external tracking (optical trackser is used to verify inside-out tracking): translation (max: $21.82 \pm 2.33$ mm), rotation: max. $8.10 \pm 2.89^{\circ}$   \\

\cite{galati2020experimental}
&
PM: 3D patient anatomy
&
Quan: PS: open abdomen surgeries
&
N/A \\

\cite{viehofer2020augmented}
&
PM: 3D foot 
&
Quan: PE: distal osteotomy
&
Mean deviation between osteotomy plane and target plane perpendicular to the second
metatarsal (anterior direction): 1.) Experienced surgeons: $4.9 \pm 4.2^{\circ}$, 2.) less experienced surgeons: $6.4 \pm 3.5^{\circ}$ \\

\cite{dennler2020augmented}
&
PM: vertebral body
&
Quan: PE: drilling pilot holes in lumbar vertebra sawbones models
&
average minimal distance of the drill axis to the pedicle wall (MAPW): 1.) Expert surgeons: $5.0 \pm 1.4$ mm, novice surgeons: $4.2 \pm 1.8$ mm\\

\cite{kriechling2020augmented}
&
PM: planned drill trajectory, IM: current drill trajectory, deviation in degrees and millimeters
&
Quan: PE: Using 3D printed scapula based on scans of human cadavers: Guidewire positioning of the central back of he 
&
mean deviation of placed guidewires from the planned trajectory: $ 2.7\pm 1.3^{\circ}$, mean deviation to the planned entry point of the placed guidewires: $2.3 \pm 1.1$ mm \\

\cite{zorzal2020laparoscopy}
&
IV: 2D plane with laparoscopic video feed, PI: 2D plane with MR image sclices
&
Qual: \& Quan: SE: laparoscopic training simulator incl. MR images and pre-recorded laparoscopic video feed 
&
N/A \\

\cite{cartucho2020multimodal}
&
PM: 3D organ models (brain and liver), PI: 2D planes with volumetric MRI/CT imaging data (with scrolling bar), 2D plane with intraoperative data (pCLE, iUS) (with transparency adjustment up and down arrows))
&
Quan: SSE: interaction with the visualisation components and exploration of holographic functionalities
&
N/A \\

\cite{rojas2020evaluation}
&
PM: 3D annotations (incision lines) and 3D surgical tools
&
Quan: SE: performing cricothyroidotomies in a simulated austere scenario (smoke and loud noises of gunshots and explosions)
&
N/A  \\

\cite{scherl2020augmented}
&
PM: 3D mandible, parotid, tumor, head, grey circles, operating menu (buttons), PI: 2D MRI images
&
Quan: PS: live parotid surgery: study persons who did not participate in the actual surgery performed manual ologram 
&
Manual Registration accuracy using fiduical markers on the head: outer borders of face: $10.09 \pm 4.23 $mm, arotid: $ 13.39 \pm 4.71$ mm, Tumor: $13.29 \pm 5.66$ mm  \\

\cite{creighton2020early}
&
PM: 3D skull and temporal bone
&
Quan: PE: Evaluation of manual target registration error using skull model
&
Target registration error: $10.62 \pm 5.90$ mm10.62  \\

\cite{jiang2020hololens}
&
PM: 3D vascular map, surrounding soft tissues, marker
&
Quan: PE: Precision verification of the vascular localization system
&
mean errors (under different conditions): min: $ 1.35 \pm 0.43$ mm , max: $ 3.18 \pm 1.32$ \\

\cite{sun2020fast}
&
PM: 3D planned mandibular reconstruction result
&
Quan: PE: 1.) Accuracy validation experiment for OST-HMD
calibration (3D printed skull with fiducials), 2.) Calibration method testing, 3.) PS: mandibular reconstruction
&
1.) Avg. root-mean-square error of control
points between rendered object and skull model: $1.30\pm0.39$  \\  

\bottomrule
\end{tabular}
\end{scriptsize}
\end{threeparttable}
\end{table*}

\end{landscape}
\pagestyle{headings}
\twocolumn

\end{appendices}
\end{document}